\documentclass[12pt, a4paper]{article}

\usepackage[toc,page]{appendix}
\usepackage{amsmath}
\usepackage{amsfonts}
\usepackage{amssymb}
\usepackage{graphicx, rotating}
\usepackage{epstopdf}
\usepackage{epsfig}
\usepackage{braket}
\usepackage{simplewick}
\usepackage{bbm}
\usepackage{empheq}
\usepackage{latexsym}
\usepackage{graphicx}
\usepackage{leftidx}
\usepackage{subcaption}
\usepackage{color}
\usepackage{amsmath,amssymb}
\usepackage{cite}
\usepackage{slashed}
\usepackage{hyperref}
\usepackage{xcolor}
\hypersetup{colorlinks, citecolor=bluscuro, linkcolor=black, urlcolor=bluscuro}
\definecolor{rossos}{cmyk}{0,1,1,0.55}
\definecolor{bluscuro}{rgb}{0.15, 0.2, .85}
\definecolor{bluchiaro}{cmyk}{1,.3,0.,0.1}

\numberwithin{equation}{section}

\newcommand{\eq}[1]{Eq.~(\ref{#1})}
\newcommand{\lag}{\mathcal{L}}

\newcommand{\nn}{\nonumber}

\newcommand{\be}{\begin{equation}}
\newcommand{\ee}{\end{equation}}
\newcommand{\bea}{\begin{eqnarray}}
\newcommand{\eea}{\end{eqnarray}}
\newcommand{\bc}{\begin{center}}
\newcommand{\ec}{\end{center}}

\newcommand{\lsc}[1]{\Lambda^\text{sc}_{#1}}

\newcommand{\beq}{\begin{equation}}
\newcommand{\eeq}{\end{equation}}
\newcommand{\bag}{\begin{align}}
\newcommand{\eag}{\end{align}}

\begin{document}
\begin{center}

\begin{flushright}
{\small Saclay-t19/022\\
SISSA 09/2019/FISI}
\end{flushright}

\vspace*{15mm}

\vspace{1cm}
{\Large \bf 
Massive Higher Spins: \\ \vspace{0.2cm}Effective Theory and Consistency}
\vspace{1.4cm}\\
{Brando Bellazzini$\,^{1}$, Francesco~Riva$\,^{2}$, Javi Serra$\,^{3}$ and Francesco Sgarlata$\,^{4}$}

 \vspace*{.5cm} 
\begin{footnotesize}
\begin{it}
 $^1$ Institut de Physique Th\'eorique, Universit\'e Paris Saclay, CEA, CNRS, F-91191 Gif-sur-Yvette, France\\
$^2$ D\'epartment de Physique Th\'eorique, Universit\'e de Gen\`eve,
24 quai Ernest-Ansermet, 1211 Gen\`eve 4, Switzerland \\
$^3$ Physik-Department, Technische Universit at M\"unchen, 85748 Garching, Germany \\
$^4$  SISSA and INFN Trieste, via Bonomea 265, 34136, Trieste, Italy
\end{it}
\end{footnotesize}

\vspace*{.2cm} 

\vspace*{10mm} 
\begin{abstract}\noindent\normalsize
We construct the effective field theory for a single massive higher-spin particle in flat spacetime. Positivity bounds of the S-matrix force the cutoff of the theory to be well below the naive strong-coupling scale, forbid any potential and make therefore higher-derivative operators important even at low energy. As interesting application, we discuss in detail the massive spin-3 theory and show that an extended Galileon-like symmetry of the longitudinal modes, even with spin, emerges at high energy.

\end{abstract}
\end{center}
\newpage

\newpage
\section*{Introduction}

Massive Higher-Spin (HS) resonances exist: atoms in electromagnetism and resonances in QCD are  familiar examples. 
The importance of HS modes  extends  potentially beyond these, into the realm of physics Beyond the Standard Model (BSM).
String excitations and large-$N$ Yang-Mills theory contain HS modes, and it is plausible that these might populate the universe at very short distances.
Such HS resonances have been shown to have distinctive phenomenological signatures in cosmology \cite{Kehagias:2017cym,Franciolini:2018eno,Bordin:2018pca,Arkani-Hamed:2018kmz}  and it is interesting to speculate about their implications for other experiments, such as in colliders or dark matter searches.

In the first part of this article, Section~\ref{sec:eft}, we discuss the effective field theory (EFT) of an isolated massive (integer) HS resonance from a particle physics perspective, in which we identify the different sectors associated, respectively, with the transverse (massless) polarizations of the HS mode, and the longitudinal ones as would-be-eaten Goldstones. Such a separation has been proven very important in the context of spin-1 particles, in the form of the equivalence theorem~\cite{Cornwall:1974km}, and for spin-2 particles \cite{ArkaniHamed:2002sp}, providing   a systematic approach to power-count the interactions and estimate the cutoff of the EFT.
Here, it will enable us to build a consistent  interacting HS EFT and to study systematically its high-energy regime.

Massless HS particles are hunted by a series of ``No-Go'' theorems, see e.g.~\cite{Bekaert:2010hw,Rahman:2013sta} for reviews. These do not represent a fundamental obstacle in the construction of a consistent massive HS EFT, but rather single out very specific low-energy structures amenable to more detailed and quantitative studies, as those we propose in this article.
The Coleman-Mandula theorem \cite{Coleman:1967ad} forbids the existence of HS conserved charges that commute with the S-matrix: the symmetry generated by such a current must be spontaneously broken. Requiring that the associated Goldstone bosons generate self-consistently a mass gap via the Higgs mechanism dictates completely the IR coupling of transverse modes and Goldstone bosons. 
Analogously, interactions which are not proportional to the mass vanish in the limit of small momentum, meaning derivative couplings that do not transmit long-range forces, in harmony with the Weinberg soft theorems \cite{Weinberg:1964ew}. 
The first part of this work is dedicated to showing that seemingly consistent low-energy interacting EFTs for HS are in fact possible.  

Yet, in QCD, the HS arise as relativistic strongly-coupled bound states of quarks and gluons, with a mass comparable to or larger than their inverse typical size, set by the interaction scale. Similarly in electromagnetism. In perturbative string theory, infinitely many HS come in towers with no parametric mass separation.
 So, in these examples, HS excitations cannot be considered in isolation: they are always accompanied by other resonances. In the language of EFT, this implies that the cutoff of an HS is of order its mass.

In the second part of this article, Section~\ref{sec:BPBs}, we focus on the question of  whether the absence of a separation of scales between mass and cutoff is a fundamental feature of HS theories or just an accident of the limited examples that have been experienced. The relevant parameter to approach this question is
\begin{equation}\label{defepsilon}
\nonumber
\epsilon=\frac{m}{\Lambda}\,,
\end{equation}
the scale $\Lambda$ being the physical cutoff.  A small $\epsilon\ll1$ implies a large range of validity for the HS EFT, while for $\epsilon\to 1$ this range shrinks to none.
Ref.~\cite{Porrati:2008ha} found that if an HS particle of spin $s$  couples to electromagnetism with charge $q$, then $\epsilon\gtrsim q^{1/(2s-1)}$ assuming the cutoff lies below the strong-coupling scale, hence implying that $\epsilon\to 1$ as the spin increases. This bound is evaded in models without  minimal coupling to photons, for instance  when a single neutral HS is at the bottom of the spectrum.  A similar bound is expected to hold for coupling HS to gravity with the replacement $q\rightarrow m/m_\text{Pl}$, and indeed Ref.~\cite{Bonifacio:2018aon} has explicitly shown that $\epsilon\gtrsim \left(m/m_{\textrm{Pl}}\right)^{1/3}$ for $s=2$.
Other consistency conditions that rely on probing the HS sector by scattering scalar particles that exchange an intermediate HS at tree level are discussed e.g.~in Ref.~\cite{Caron-Huot:2016icg}. Generalizing the causality constraints of  Ref.~\cite{Camanho:2014apa}, the positivity of the eikonal phase shift in the tree-level scattering of an HS gravitationally coupled  to a scalar, sets other bounds under certain assumptions~~\cite{Afkhami-Jeddi:2018apj}. While these gravitational bounds are robust--gravity is universally coupled--they are not directly relevant for phenomenological purposes far below the Planck scale--gravity is very weakly coupled at low energy--and in fact these constraints evaporate as $\Lambda/m_{\textrm{Pl}}\rightarrow 0$ or when new, light, degrees of freedom are more important than gravity. 
  
In this article, we propose a new class of constraints on $\epsilon$, which do not rely on coupling to external probes, but rather target directly the consistency of the self-interacting HS theory, based on our construction and understanding developed in Section~\ref{sec:eft}.
We first discuss the simple requirement that the EFT be perturbative in its range of validity. Then the constraints from perturbativity are superseded in theories where the putative microscopic theory from which the HS EFT emerges is Lorentz invariant, local and unitary. Dispersion relations for forward elastic scattering amplitudes lead indeed to
certain positivity bounds \cite{Adams:2006sv,Bellazzini:2016xrt,Bellazzini:2017fep} that lower the cutoff to be parametrically close to the HS mass. More specifically, for a generic potential $\lambda_L \Phi^4$ the (beyond) positivity bounds for the longitudinal polarizations give $\epsilon\gtrsim (\lambda_L/16\pi^2)^{1/(8s-4)}$, which is more stringent than the perturbativity bound set by the strong-coupling scale. The bound implies that the higher the spin the smaller the gap, $\epsilon\to1$, unless the coupling is simultaneously taken smaller.   A similar bound holds as well for special types of potentials that display higher strong-coupling scales analogously to the case of $\Lambda_3$-theory of massive gravity \cite{ArkaniHamed:2002sp,deRham:2010kj}: the associated beyond-positivity bound for the longitudinal polarizations remains much more stringent, alike the case for the massive spin-2 theory \cite{Bellazzini:2017fep}.  We remark that the beyond-positivity bounds constrain as well the cutoff of the transverse modes, see Eq.~(\ref{ttttbound}).  
Finally, if certain conditions about weak coupling are met, we find that $\epsilon\to 1$ independently on the value of $s>2$.
\newpage

\section{Effective theory of massive Higher Spins}\label{sec:eft}

A free \emph{massive} spin-$s$ particle can be described by a field $\Phi$ transforming in the representation $D(s/2,s/2)$ of the Lorentz group and satisfying on-shell the usual Klein-Gordon equation, together with the traceless and transverse conditions,\footnote{{\bf Notation:} 
Given a rank-$s$ totally symmetric field $\phi_{\mu_1...\mu_s}$, we use the following notation
\begin{equation}\label{EoMHS}
\phi = \phi_{\mu_1...\mu_s}\,,\qquad\phi^\prime = \phi^\prime_{\mu_3...\mu_s} = \eta^{\mu_1\mu_2}\phi_{\mu_1...\mu_s}\,,\qquad
\partial \phi = \partial_{(\mu}\phi_{\mu_1...\mu_s)}\,,\qquad \partial \cdot \phi = \partial^\alpha\phi_{\alpha\mu_2...\mu_s}\,,
\end{equation}
where (anti) symmetrizations are defined without normalization factors, e.g.~$a_{(\mu} b_{\nu)} = a_{\mu}b_{\nu}+b_{\nu}a_{\mu}$ and $a_{[\mu} b_{\nu]} = a_{\mu}b_{\nu}-b_{\nu}a_{\mu}$. 
We use mostly plus signature $\eta_{\mu\nu} = \text{diag}\left(-,+,+,+\right)$. The $\phi^T$ represents the traceless part of $\phi$, namely $\phi^T = \phi-\frac{1}{2s}\eta\phi^\prime$.}
\begin{equation}
\label{EOMmassivehs}
\left(\square - m^2\right)\Phi = 0\,,\qquad \Phi^\prime = 0\,,\qquad \partial \cdot \Phi = 0\,.
\end{equation}
More precisely, the free field can be constructed in terms of the physical polarizations 
$$\braket{0|\Phi_{\mu_1...\mu_s}(0)|\textbf{p},\sigma} = \epsilon_{\mu_1...\mu_s}(\textbf{p},\sigma)\,,$$
which satisfy the on-shell conditions, with $\sigma$ labelling the spin-$z$ component. 
At high energies, $E^2 \gg m^2$, the solutions to the equations of motions \eq{EOMmassivehs} are defined up to gauge transformations 
\begin{equation}
\label{gaugeSymmetryPolarizations}
\epsilon_{\mu_1...\mu_s} \rightarrow \epsilon_{\mu_1...\mu_s}  + p_{(\mu_1} \chi_{\mu_2...\mu_s)}\,,
\end{equation}
parametrized by a transverse ($p\cdot \chi =0$) and traceless ($\chi^\prime =0$) tensor $\chi$, which transforms as a lower-spin polarization. The $\chi$ represent the  longitudinal modes of the massive multiplet in the high-energy regime. 
Therefore, in the EFT perspective, theories of interacting massive HS particles of integer spin can be equivalently separated into the EFTs for the transverse and for the longitudinal modes, the latter corresponding to lower-spin would-be Goldstone bosons, also known as Stueckelberg fields.  

\subsection{The  Transverse Sector} 

The transverse sector contains, in isolation, a \emph{massless} spin-$s$ state i.e.~two degrees of freedom. In order to extend the description off-shell with a Lagrangian, it is useful to relax the on-shell conditions  $p \cdot \chi =0$ while enlarging the gauge symmetry group Eq.~(\ref{gaugeSymmetryPolarizations}). We introduce traces $\Phi^\prime$ as  pure gauge degrees of freedom  while keep working with double-traceless fields $\Phi^{\prime\prime}=0$.\footnote{This is  analogous to the case of a massless spin-2 in General Relativity, where one can choose to work with a traceful $h_{\mu\nu}$ by enlarging the volume-preserving gauge transformations to generic diffeomorphisms. Going back to a traceless $h_{\mu\nu}$ is just a question of gauge fixing.}
The massless spin-$s$ field enjoys then the gauge invariance\footnote{We are not aware of any consistent non-abelian extensions of the gauge transformation in flat space-time, therefore we focus on abelian transformations. }
\begin{equation}
\label{gaugeSymmetryHS}
\Phi \rightarrow \Phi + \partial \xi\,,\qquad \xi^\prime =0\,,
\end{equation}
which makes only two components of the field  propagate (see \cite{Rahman:2013sta} for a pedagogical review). The quadratic gauge invariant Lagrangian is \cite{Fronsdal:1978rb}
\begin{align}
\label{masslessHSLagrangian}
\begin{split}
\mathcal{L}_{s} &=\frac{s}{2}\left(\partial\cdot \Phi \right)^2 -\frac{1}{2}\left(\partial_\mu \Phi\right)^2  + \frac{s(s-1)}{2}\left[\Phi^\prime\cdot \partial\cdot \partial\cdot \Phi+\frac{1}{2}\left(\partial_\mu \Phi^\prime\right)^2 + \frac{(s-2)}{4}\left(\partial \cdot \Phi^\prime \right)^2\right]\,,
\end{split}
\end{align}
from which the field equations can be collected in terms of the so-called Fronsdal tensor
\begin{equation}
\label{FronsdalTensor}
\Gamma_s \equiv \square \Phi -\partial \partial \cdot \Phi + \partial \partial \Phi^\prime = 0 \,,
\end{equation}
that can be used to write the kinetic lagrangian in the more compact form
\begin{align}
\label{kinTermFronsdal}
\mathcal{L}_{s_T}^0&= \frac{1}{2}\Phi\cdot \left(\Gamma_s -\frac{1}{2}\eta \Gamma^\prime_s \right)\equiv  \Phi \cdot  \hat{\Gamma}_s\,.
\end{align}

\paragraph{Interactions.} 
Self interactions of massless HS can be written in terms of $\Gamma_s$ in \eq{FronsdalTensor} and the generalized Riemann tensor
\begin{equation}
\bcontraction[4pt]{R_{\alpha_1\alpha_2...\alpha_s\mu_1\mu_2...\mu_s} = \partial\alpha_1}{{}}{_{\alpha_2...\alpha_s}\Phi_{\mu_1}}{{}\!}\,
\bcontraction[6pt]{R_{\alpha_1\alpha_2...\alpha_s\mu_1\mu_2...\mu_s} = \partial\alpha_1\alpha}{{}}{_{\alpha_3}\Phi_{\mu_1....}}{{}_{\mu_2}\!}\,
\bcontraction[8pt]{R_{\alpha_1\alpha_2...\alpha_s\mu_1\mu_2...\mu_s} = \partial\alpha_1\alpha_2}{{}_{...\alpha_3}}{\Phi_{\mu_1\mu_2}}{{}_{...\mu_s}\!}\,
\mathcal{R}_{\alpha_1\alpha_2...\alpha_s\mu_1\mu_2...\mu_s} = \partial_{\alpha_1\alpha_2...\alpha_s}\Phi_{\mu_1\mu_2...\mu_s}
\end{equation}
with anti-symmetric contractions. These generalizations of the  Christoffel symbols and curvature tensor of spin-2 fields, introduced in Ref.~\cite{deWit:1979sib}, are linear in the HS field and manifestly gauge invariant
\begin{equation}
\delta_\xi \Gamma_{\mu_1...\mu_s} = 3 \partial_{(\mu_1}\partial_{\mu_2}\xi_{\mu_3...\mu_s)\sigma}^{\phantom{\mu_3...\mu_s)\sigma}\sigma} = 0\, , \quad\quad
\delta_\xi \mathcal{R}_{\alpha_1\alpha_2...\alpha_s\mu_1\mu_2...\mu_s}= 0\,,\label{shiftriem}
\end{equation}
where the first relation holds only for traceless $\xi$ parameters, while the second involving $\mathcal{R}$ is satisfied also for $\xi_{\mu_3...\mu_s\sigma}^{\phantom{\mu_3...\mu_s\sigma}\sigma}\neq 0$.
These are the necessary ingredients to construct  gauge invariant interactions.  Interactions involving the Fronsdal tensor~$\Gamma_s$ are proportional to the equations of motion (see \eq{FronsdalTensor}) and can therefore be removed by a suitable field redefinition (the same holds for other operators with less than $s$ derivatives per field \cite{deWit:1979sib} that we do not discuss here).

For a single flavour and odd spin, cubic interactions are forbidden,\footnote{Poincar\'e symmetry implies that the amplitude for the state of any two spin-$s$ particles, with $s$ odd, from the cubic vertex be anti-symmetric (e.g. \cite{Elvang:2013cua}). Moreover, two of the helicities in a cubic vertex with spin $s$ are always equal; therefore it can be non-zero only in the presence of a non-trivial flavour structure, in which case they may be constrained by the arguments of Ref.~\cite{Hinterbichler:2017qcl}}
while for a non-trivial flavour structure or even spin, their contributions to scattering amplitudes is always smaller than those from the quartic contact-term, since they scale with more powers of energy.
We can therefore focus on quartic  interactions for the \textit{transverse} polarizations, schematically of the form
\begin{equation}
\label{transverseInt}
\lag_{s_T}^\text{int} = \frac{(\mathcal{R}_{\mu_1...\mu_s}^{\alpha_1...\alpha_s})^4}{f_T^{4s}}+\cdots
\end{equation}
with  ${f}_T$ a scale characterizing the interaction strength, and the dots standing for terms with higher derivatives (whose suppression scale is discussed in Section \ref{sec:BPBs}), or more insertions of $\mathcal{R}$, relevant for processes with more than 4 external states. Eq.~(\ref{transverseInt}) implicitly hides thousands of possible contractions; from studying HS scattering amplitudes it is however obvious that the number of physically independent contractions, as long as massless states are concerned, is {equivalent to the number of helicity 4-point amplitudes: only four parity-invariant combinations are independent.}

As expected, the highly irrelevant operators in \eq{transverseInt} vanish at low energy, complying with the Weinberg soft theorems. 
Moreover, since these interactions are trivially invariant under \eq{gaugeSymmetryHS}, they do not give rise to any HS charge, in agreement with the Coleman-Mandula theorem.   
Yet, the exactly massless limit is incompatible with a finite coupling to gravity \cite{Porrati:2012rd}, an argument that can be evaded only at finite mass.

\subsection{The Longitudinal Sector}

The longitudinal sector, external a priori to the transverse one, provides the missing longitudinal modes necessary to describe a \emph{massive} multiplet of spin $s$ in a somehow complicated generalization of the known case of massive vector theories.
It consists of a tower of lower-rank  double-traceless tensor fields $\phi_{(k)}$ of spin $k=s-1,s-2,\dots,0$,  transforming non-linearly under the would-be spin-$s$ gauge symmetry. In order to project out unnecessary  degrees of freedom otherwise present in this redundant description, the fields in the longitudinal sector transform under a tower of gauge transformations 
\begin{align}
\label{gaugeSymmetriesGoldstones}
\begin{split}
\delta \phi_{(s-1)} &=  m \sqrt{s}\,\xi + \partial \lambda_{(s-2)}\\
\delta \phi_{(s-2)} &= \lambda_{(s-2)} + \partial \lambda_{(s-3)}\\
&\dots\\
\delta \phi_{(0)} &= \lambda_{(0)}
\end{split}
\end{align}
where $\lambda_{(k)}$ are traceless gauge parameters, and the  fields have non-canonical mass dimension, $[\phi_{(s-k)}] =2-k$; the dimensionfull parameter $m$ will be linked later to the HS physical mass.\footnote{The normalization of the gauge parameter $\xi$ has been chosen to reproduce the quadratic Lagrangian with bare mass $m$, see Eq.~(\ref{totLagrangianHS}).}
The symmetry associated with the $\xi$ parameters will eventually be gauged when longitudinal and transverse sectors interact to concoct a massive HS state; but for the purpose of studying the longitudinal sector in isolation, $\xi$ should be thought of as the parameter of a \textit{global} symmetry. This is the HS analog of the shift symmetry characteristic of the scalar Goldstone bosons eaten into massive spin-1 states. 

Analogously, but within the longitudinal sector,  each  Goldstone boson $\phi_{s-n}$ \textit{eats} the  lower-level Goldstone boson $\phi_{s-n-1}$ whose shift symmetry $\phi_{s-n-1}\rightarrow \phi_{s-n-1}+\lambda_{s-n-1}$ has been gauged in Eq.~(\ref{gaugeSymmetriesGoldstones}). Another simple way to derive such a cascade of shift-symmetries, whose Goldstone bosons are gauged-away, is by Kaluza-Klein reduction of a massless HS in 5 dimensions \cite{Porrati:2008ha,Bianchi:2005ze}.

A Lagrangian for the longitudinal sector is easily built in terms of  the double-traceless combinations
\begin{equation}\label{defvarphi}
\varphi_{(k)}\equiv\phi_{(k)}-\partial \varphi_{(k-1)}^T \,,
\end{equation}
which, under the web of gauge transformations Eq.~(\ref{gaugeSymmetriesGoldstones}), shift simply as
\begin{equation}
\delta \varphi_{(k)} = \lambda_{(k)}\,,\qquad \delta \varphi_{(s-1)} =\lambda_{(s-1)}\equiv m\sqrt{s}\, \xi \,.
\end{equation}
Then, simple invariants can be built in terms of derivatives of $\varphi_{(s-1)}$ (given that, in isolation, $\xi$ is constant) as well as single traces $\varphi_{(k)}^\prime$ (given that the gauge parameters $\lambda_{(k)}$ are traceless).
In addition to these,  the generalized Christoffel and Riemann tensors for $\phi_{(s-1)}$ (but not the ones of lower spin)  can also be used to build invariants.
At the quadratic level the most general Lagrangian, invariant under \eq{gaugeSymmetriesGoldstones} up to total derivatives, is therefore 
\begin{equation} \label{longfree}
\mathcal{L}_{s_L}^0 = \phi_{(s-1)}\cdot \hat{\Gamma}_{s-1}+ \mathcal{L}_\text{aux} 
\end{equation}
with $\hat\Gamma_{s-1}$ defined in Eq.~(\ref{kinTermFronsdal}) and
\begin{equation}
\label{auxLagrangian}
\mathcal{L}_\text{aux} = \sum_k b_k\left(\partial_\mu \varphi_{(s-k)}^\prime\right)^2+ \tilde{b}_k\left(\partial\cdot \varphi_{(s-k)}^\prime\right)^2+c_k\left( \varphi_{(k)}^\prime\right)^2 +\sum_{k< k'} a_{k,k'}\left(\partial\cdot \varphi^\prime_{(s-k)}\right)\cdot \varphi_{(s-k')}^\prime \,,
\end{equation}
where $b_k,\tilde{b}_k,a_{k,k'},c_k$ are dimension-full coefficients. Terms with more fields, will-be interactions, can be written instead as polynomials in
\begin{equation}\label{longint}
\varphi^\prime_{(k)}\,,\quad\quad \partial \varphi_{(s-1)}\,,
\end{equation}
and their derivatives. We stress that \eq{longfree} should not be thought as a weakly coupled Lagrangian for particles in isolation, as the standard kinetic terms for the lower-spin Goldstone fields are induced only after mixing with the transverse sector.  Similarly to the situation in massive gravity, and contrary to the spin-1 case, the longitudinal sector in isolation does not describe a theory of particles. Yet, coupling it to the transverse sector, accompanied by the tuning of a finite set of parameters (analog to Fierz-Pauli tuning in massive gravity), will make the longitudinal components  sprout  to life and defer the would-be ghost instabilities beyond the cutoff, as we discuss next. 

\subsection{Mixing between Sectors, Tuning and Higgsing}
\label{subsec:mixing}

Interactions between transverse and longitudinal sectors are generically controlled by the most relevant operator: the minimal coupling of the transverse spin-$s$ fields with the current of the longitudinal sector, associated to global shifts of $\phi_{(s-1)}$ in \eq{longfree}, namely 
\begin{equation}\label{mixing}
\lag^{\text{mix}}=-\Phi\mathcal{J}\,.
\end{equation}
In practice this is equivalent to weakly gauging the global symmetry $\xi$ in \eq{gaugeSymmetriesGoldstones}, that is promoting it to a \textit{local} symmetry. 
The current, in its  expression invariant under the $\lambda_{k}$ gauge symmetries of the longitudinal sector in \eq{gaugeSymmetriesGoldstones},  is given by
\begin{equation}
\label{currentgaugeinv}
\mathcal{J} = -\frac{m}{\sqrt{s}}\left[\partial\varphi_{(s-1)} -2 \eta \partial \cdot \varphi_{(s-1)} + \frac{1}{2}\eta \partial \varphi_{(s-1)}^\prime \right].
\end{equation}
Since it is charged under the gauged $\xi$-shift,  $\delta_\xi \mathcal{J}\sim m^2$, a mass term for the spin-$s$ field is necessary in order to make the full Lagrangian $\lag_{s_T}^0+\lag_{s_L}^0+\lag^{mix}$ invariant under local $\xi$ transformations. The resulting quadratic Lagrangian is
\begin{equation}
\label{totLagrangianHS}
\mathcal{L}^0 =
\Phi \cdot \hat{\Gamma}_s +\phi_{(s-1)}\cdot \hat{\Gamma}_{s-1}-\frac{m^2}{2} \left[\Phi^2-\frac{s(s-1)}{2}{\Phi^\prime}^2\right] -\Phi \cdot \mathcal{J} +\mathcal{L}_\text{aux} \,.
\end{equation}
By a proper gauge choice--the HS analog of the  unitary gauge--both $\phi_{(0)},\phi_{(1)}$ and the traceless component of the higher-spin Golstone fields $\phi_{(k>1)}$ can be removed from \eq{totLagrangianHS}. On the other hand, the single-traces of the Goldstones fields with $k=2,\dots,s-1$ cannot be removed and  appear as auxiliary fields of spin $0,1,\dots,s-3$:
\begin{equation}
\label{unitaryGauge}
\phi_{(k)} = \frac{1}{2k} \eta \phi_{(k)}^\prime\,.
\end{equation}


\paragraph{Tuning conditions.} 
For generic values of the coefficients in Eq.~(\ref{auxLagrangian}), {single-traces are dynamical and ghost-like (e.g.~double poles of propagators at low energy). However, there exists a specific choice which fixes the number of degrees of freedom to $N_\text{dof} = 2s-1$, removes ghost-like instabilities, and makes these fields \textit{auxiliary}, in the sense that the equations of motions are algebraic, $\phi_{(k)}^\prime = 0$.}
We can find this choice by demanding that ghost-like kinetic terms for the Goldstone fields be absent (for instance, the Fierz-Pauli tuning of a massive spin-2 field theory projects out the term $(\square \phi_{(0)})^2$).\footnote{Ref.~\cite{Singh:1974qz} obtains the same result  by enforcing  the equations of motion Eqs.~(\ref{EOMmassivehs},\ref{unitaryGauge}).}

We single out a piece $\mathcal{I}_{s-2}$ from the current Eq.~(\ref{currentgaugeinv}),
\begin{equation}
\label{defIs2}
\mathcal{J} = \tilde{\mathcal{J}} + \mathcal{I}_{s-2}\,,\qquad \mathcal{I}_{s-2} = \frac{m}{\sqrt{s}}\left[2\partial \partial \varphi_{(s-2)}^T-2\eta\square \varphi_{(s-2)}^T-\eta \partial \partial\cdot\varphi_{(s-2)}^T\right]\,,
\end{equation}
where we recall that $\varphi_{(s-2)}^T$ is the traceless part of $\varphi_{(s-2)}$.  
A standard kinetic term  for $\phi_{(s-2)}$, i.e.~of the form of \eq{masslessHSLagrangian}, can be induced under the field redefinition
\begin{equation}
\label{fieldRedefSpinS}
\Phi \rightarrow \Phi + \kappa\, \eta \varphi_{(s-2)}^T\,, \quad\quad \kappa = \frac{m}{\sqrt{s}(s-1)}\,,
\end{equation}
where $\kappa$ has been chosen to cancel the kinetic term transformation under \eq{fieldRedefSpinS},
\begin{equation}
\delta \left(\Phi \cdot \hat{\Gamma}_s\right) =  2\Phi \cdot \delta\hat{\Gamma}_s+\kappa\,\eta\varphi_{(s-2)}^T\cdot  \delta\hat{\Gamma}_s\,,\qquad \delta \hat{\Gamma}_s = \kappa \frac{\sqrt{s}}{2m}(s-1) \mathcal{I}_{s-2}\,,
\end{equation}
against  the mixing $-\Phi\cdot \mathcal{I}_{s-2}$. 
This generates the standard kinetic term of a traceless field\footnote{ With $\Gamma\left(\varphi^T\right)$ we mean the Fronsdal tensor \eq{FronsdalTensor} projected on the traceless components of $\varphi$.}
\begin{align}\label{kintermvarphss2}
\frac{m^2}{2}(2s-1) \varphi_{(s-2)}^T \cdot \Gamma(\varphi^T_{(s-2)}) -\frac{3}{2}m^2(s-1)(s-2)\left(\partial \cdot \varphi_{(s-2)}^T\right)^2\,, 
\end{align}
with an additional piece that is canceled by tuning the coefficient 
\begin{equation}
c_{s-1}=\frac{3}{8}m^2(s-1)(s-2)\,.
\end{equation}
%
 From \eq{kintermvarphss2} and the definition of $\varphi$ in \eq{defvarphi}, we recognise that the Lagrangian contains still  ghost-like terms $\sim \left(\partial^2 \varphi_{(s-3)}^T\right)^2$ for the spin-$(s-3)$ field. 
With a proper tuning of the others coefficients in Eq.~(\ref{auxLagrangian}), we can generate a gauge-invariant kinetic term for the traceful field $\varphi_{s-2}$ such that these ghost-like kinetic terms cancel as well, and we can repeat the procedure until all ghost-like kinetic terms are absent. We detail this for the $(s-3)$-Goldstone field in Appendix~\ref{appendix:tuning}.

The resulting Lagrangian describes the $2s-1$ degrees of freedom of a massive HS state, with mass $m$ \cite{Singh:1974qz}.


\subsection{Interactions}

We have  identified different types of interactions: those that originate in the transverse sector in terms of the Riemann tensor \eq{transverseInt},  and those from the longitudinal sector originally built with the building blocks in \eq{longint}: $\partial \varphi_{(s-1)}$ and $\varphi^\prime_{(k)}$.
Operators involving the auxiliary fields $\varphi^\prime_{(k)}$, to which no physical poles are associated, may be removed by means of field redefinitions.\footnote{For instance, we can iteratively remove interaction of the form $\phi^\prime_{(s-1)} G \left[  \partial^n, \phi_{(k)}^\prime, \Phi  \right]$ with $G[...]$  a polynomial in fields with at most $n$ derivatives, by the variation of the mass term of $\phi_{(s-1)}^\prime$ under the redefinition $\phi_{(s-1)}^\prime \rightarrow \phi_{(s-1)}^\prime - G \left[  \partial^n, \phi_{(k)}^\prime, \Phi  \right]/(2 c_{s-1}).$
}
Since interactions built with $\partial \varphi_{(s-1)}$ are not invariant  under the local transformations associated with $\xi(x)$, one should promote the  symmetrized derivatives of $\varphi_{(s-1)}$,  gauged by the $\Phi$-field, to covariant derivatives\footnote{Anti-symmetric combinations, e.g.~$\partial_{[\alpha}\varphi_{\mu_1]\mu_2...\mu_s}$, could be gauged by other fields as in Eq.~(\ref{generalCovD}), however by assumption these do not populate the infrared physics. Therefore, we omit these combinations in the following, noticing that they cannot be dynamically generated as long as the interactions only involve symmetrized derivatives.}
\begin{equation}
\label{generalCovD}
D\varphi_{(s-1)}\equiv \partial\varphi_{(s-1)}-m \sqrt{s}\,\Phi\,. 
\end{equation} 
For instance, an interaction $\left(\partial \varphi_{(s-1)}\right)^4/f_L^4$, with $f_L$ the scale controlling its strength in the longitudinal sector in isolation, is written as $\left(D\varphi_{(s-1)}\right)^4$ and can be read in the unitary gauge schematically as
\begin{equation}
\lambda_L (\Phi_{\mu_1...\mu_s})^4\,,\qquad \lambda_L\propto m^4/f_L^4\,. 
\end{equation}
Notice that in non-abelian theories, like massive gravity, longitudinal interactions arise already from the mass term, so that their coupling \emph{equals} $m^2$; for abelian theories instead, mass and couplings are independent and therefore $\lambda_L\sim m^4$ remains a conservative estimate. 
Of course, one can always add more irrelevant operators in each sector by including more derivatives, see e.g.~Eq.~(\ref{generalLagrangian}), or consider mixed-type interactions involving Riemann tensors, e.g.~$\frac{m^2}{f_L^2 f_T^6} \Phi^2 \mathcal{R}^2\, $.

\subsection{Spin-3 and decoupling limit}
\label{sec:spin3}

The explicit example of a massive spin-3 will make more concrete the points introduced above.
To make the notation clearer, we label the Goldstone fields as 
$$\phi_{(2)} \equiv H_{\mu\nu}\,,\,\,\,\phi_{(1)} \equiv A_\mu\,, \,\,\,\phi_{(0)}\equiv \pi \,. $$
In the unitary gauge, only the trace $H=\phi^\prime_{(2)}$ survives as an auxiliary field and the free Lagrangian Eq.~(\ref{totLagrangianHS}) becomes
\begin{align}
\label{freeLagrangianUnitaryGauge}
\begin{split}
\mathcal{L}^{0} &=-\frac{1}{2}\left(\partial_\sigma\Phi_{\mu\nu\rho}\right)^2+\frac{3}{2}\left(\partial_\mu \Phi^{\mu\nu\rho}\right)^2 + \frac{3}{4}\left(\partial_\mu \Phi^\mu\right)^2+\frac{3}{2}\left(\partial_\mu \Phi_\nu\right)^2 + 3\Phi_\rho\partial_\mu\partial_\nu \Phi^{\mu\nu\rho}\\
& -\frac{m^2}{2}\left[\Phi_{\mu\nu\rho}^2-3\Phi_\mu^2\right]+\frac{3}{16}\left(\partial H\right)^2 + \frac{3m^2}{4} H^2  + \frac{\sqrt{3}}{4}m \Phi_\mu \partial^\mu H\,,
\end{split}
\end{align}
with equations of motion
\begin{equation}
\label{EOMspin3}
\left(\square - m^2\right)\Phi_{\mu\nu\rho}=0\,,\qquad H=0\,,\qquad \partial_\mu \Phi^{\mu\nu\rho} = 0\,,\qquad \Phi_\mu = 0\,.
\end{equation}
The most relevant self-interactions of each separate type are\footnote{We remark that for odd spins a cubic potential is not allowed by Lorentz invariance, while for even spins it is.
 For simplicity, we focus in what follows on situations symmetric under $\Phi\rightarrow -\Phi$ and comment in the conclusions about the possible impact of trilinear couplings like $\Phi^3$, $\Phi^2\partial\Phi$, \dots. } 
\be
\label{fullPowerCounting}
\mathcal{L}_\text{int} =
 -\lambda_L \Phi^4 + \frac{\mathcal{R}^4}{f_T^{12}} + \cdots \,.
\ee
While there are thousands possible contractions of four Riemann tensors, only three independent contractions $\Phi^4$ give non-vanishing contributions to on-shell scattering amplitudes,
\begin{equation}
\label{Potential}
-\lambda_L \Phi^4  = \lambda_1\, \Phi_{a}^{\phantom{a}de}\Phi^{abc}\Phi_{bd}^{\phantom{bd}f}\Phi_{cef} + \lambda_2\, \Phi_{ab}^{\phantom{ab}d}\Phi^{abc}\Phi_c^{\phantom{c}ef}\Phi_{def}+\lambda_3\left(\Phi_{abc}\Phi^{abc}\right)^2=-V(\Phi)\,,
\end{equation}
and constitute a potential for $\Phi$.

\paragraph{High-energy limit.}
The high-energy regime  $E\gg m$ can be understood by studying the behaviour of scattering amplitudes in terms of the eaten Goldstone bosons with lower spins. 
The procedure outlined above, and detailed in Appendix~\ref{appendix:tuning}, aimed at finding a ghost-free quadratic HS EFT, delivers as a by-product the high-energy theory in which polarizations of different helicities behave as independent massless states of spin $(s-k)$, $k=0,\dots,s$.

The mixings in \eq{freeLagrangianUnitaryGauge} are resolved in a way that keeps the $m\to 0$ limit manifestly smooth,
\begin{align}
\label{fieldRedef}
\begin{split}
\Phi_{\mu\nu\rho}&\rightarrow \Phi_{\mu\nu\rho} +\frac{m}{2\sqrt{3}}\left[\, \eta_{(\mu\nu}A_{\rho)}-\eta_{(\mu\nu}\partial_{\rho)} \pi\right] \,,\\
H_{\mu\nu}&\rightarrow H_{\mu\nu} +\frac{5}{2}m^2\,\eta_{\mu\nu} \pi \,,
\end{split}
\end{align}
and isolates the propagating ``high-energy'' degrees of freedom as massless spin-3, 2, 1 and 0 states, associated with the fields $\Phi,H,A,\pi$ in the Lagrangian 
\begin{equation}
\label{decouplingL}
\mathcal{L}^{0}_{m\to0} =  \Phi_{\mu\nu\rho}\hat{\Gamma}_3^{\mu\nu\rho} +H_{\mu\nu}\hat{\Gamma}_2^{\mu\nu}  -\frac{5}{4}{\hat F}_{\mu\nu}^2 -15(\partial\hat \pi)^2\,,
\end{equation}
with $A_\mu = \hat{A}_\mu/m$ and $\pi= \hat{\pi}/m^2$  the (almost) canonically normalized fields; see Appendix~\ref{appendix:decouplingLimit} for expressions away from the massless limit. This is the analog of \eq{kintermvarphss2}, when all the ghost-like kinetic terms have been removed, and it corresponds  to the HS-equivalence theorem for massive HS states: at high energy their dynamics separates into that of transverse modes and the longitudinal ones. 

In the massless limit, the quadratic action is trivially invariant under the $N$-th order polynomial symmetries
\begin{align}
\label{ExtShift}
\hat \pi&\rightarrow \hat \pi + f^{(N)}(x)\,,\qquad \hat A_\nu \rightarrow \hat A_\nu + C_\nu^{(N)}(x)\,,
\end{align}
with 
\begin{align}
\label{LevelN}
f^{(N)}(x) &=  \sum_{n=0}^N \frac{1}{n!}c^T_{\mu_1...\mu_n}x^{\mu_1}\ldots x^{\mu_n}\,,\\
C_\nu^{(N)}(x) &= \sum_{n=1}^N \frac{1}{n!}b^T_{\nu \mu_1...\mu_n}x^{\mu_1}\ldots x^{\mu_n}\,
\end{align}
where $c^T_{\mu_1...\mu_n}$ and $b^T_{\nu \mu_1...\mu_n}$ are traceless tensors and symmetric under $\nu_i\leftrightarrow \nu_j$ \footnote{We thank David Stefanyszyn for useful discussions.}. Generically, the latter corresponds to a generalization of the Galileon symmetry to HS fields \cite{futurework} but also includes gauge symmetries, i.e. $C_{\nu}^{(N)} = \partial_\nu \Omega$, when all the indexes are totally symmetrized. 	For instance, $N=1$ non-gauge transformations are of the kind $b^T_{[\mu\nu_1]}$. Notice that these transformations are true symmetries (not gauge redundancies), as they act on the transverse modes of the vector field. The scalar transformation is instead an extended shift-symmetry of the type described in Ref.~\cite{Hinterbichler:2014cwa}. Generically, the interactions are expected to spoil this invariance (for any $N$) and therefore it is interesting to probe the set of operators which preserves the highest number of symmetries, at least in the high-energy regime.
Interactions made of $\partial^{N+1}\hat{\pi}$ and $\partial^{N+1}\hat{A}_\mu$ are trivially invariant for any $N$; we are interested instead in non-trivial invariants with less then $N+1$ derivatives per field, of which we provide  a novel example in \eq{tunedInteractions}.

Interactions from the longitudinal sector (in general interactions that cannot be written in terms of $\mathcal{R}$) are rewritable in terms of the covariant derivative 
\begin{align}
\label{HcovDev2}
D_{(\mu}\varphi_{(2)\nu\rho)} &\equiv \partial_{(\mu}H_{\nu\rho)} -\frac{2}{m}\partial_{(\mu\nu}\hat{A}_{\rho)}+\frac{6}{m^2}\partial_{\mu\nu\rho}\hat{\pi}
+3\eta_{(\mu\nu}\partial_{\rho)}\hat{\pi}-\frac{1}{2}m \, \eta_{(\mu\nu}\hat{A}_{\rho)}- \sqrt{3}m\Phi_{\mu\nu\rho}\,.
\end{align}
This makes it clear that the polarisation vectors of spin 3, 2, 1, and 0 grow respectively as $E^0,E^1,E^2,E^3$ with $E$ the particle energy, and that in the high-energy regime we generically expect $N=2$ symmetry for the scalar mode and $N=1$ for the vector. 

Since the powers of energy are accompanied by  inverse powers of mass,  and $E/m \gg 1$, they lead in practice to a premature loss of predictivity, or in other words to a low strong-coupling scale, as we discuss next.

\section{Structural Constraints}
\label{sec:BPBs}

There are a number of reasons why  the ratio between mass and cutoff, $\epsilon$, cannot take arbitrary values in a HS theory with a given interaction.  First, analogously to massive gravity, the theory becomes strongly coupled  at energy scales $\Lambda^\text{sc}$ parametrically close to the particle's mass, leading to a constraint on $\epsilon$ if one demands an energy range of calculability.  Second, dispersion relations for forward scattering amplitudes imply UV-IR relations~\cite{Adams:2006sv,Bellazzini:2016xrt}  when the S-matrix is unitary, analytic, crossing symmetric, and polynomially bounded in the forward limit (the latter condition is implied by the Froissart bound \cite{Froissart:1961ux,Martin:1965jj} in local UV completions). These  lead to different classes of positivity constraints on the parameter $\epsilon$ that we are now going to study. Non-forward dispersion relations may also be exploited in weakly coupled theories~\cite{deRham:2017zjm,deRham:2018qqo}; we leave the exploration of those constraints to future work. 

\subsection{Strong Coupling}

We focus on $2\to2$ scatterings, whose amplitudes $\mathcal{M}$ have dimension of a coupling-squared, so that we define 
\begin{equation}\label{couplingdefinition}
g^2(E)\equiv \mathcal{M} (E)
\end{equation}
with $E=\sqrt{s}$ the center of mass energy. Different processes and interactions can be associated with different coupling strengths, for each of them we define the value of the coupling at the physical cutoff $\Lambda$ as~$g^2\equiv g^2(\Lambda)$. The EFT becomes strongly coupled at $E=\Lambda^\text{sc}$ when 
\begin{equation}\label{scdef}
g^2(\Lambda^\text{sc})=\mathcal{M}(\Lambda^\text{sc})\simeq (4\pi)^2\,. 
\end{equation}
Therefore the \emph{strong-coupling scale} $\Lambda^\text{sc}$ corresponds roughly to the largest possible value for the  physical cutoff,~$\Lambda<\Lambda^\text{sc}$, for a useful calculable EFT: it does not \emph{necessarily} correspond to the physical mass of a particle $\Lambda$, but it is the ultimate energy above which the theory changes regime and a new EFT description is required. 
This is the analog of the strong-coupling scale $4\pi m_W/g$ in the scattering of longitudinally polarised $W$ bosons, for which a SM description without the physical Higgs boson ceases to make sense, but lies much above both $m_W$ and the Higgs mass. 

For concreteness we discuss spin-3 particles and  focus first on interactions of the simple form $\mathcal{R}^4$ and $\Phi^4$. 
Amplitudes for the scattering of different helicities  (which we label  $\sigma=T,T',H,H',V,V',S$ for spin 3, 2, 1 and 0 respectively) 
exhibit different rates of energy-growth, some of which are illustrated in Table~\ref{tab:Scalings}. 
\begin{table}[t]
\center
\begin{tabular}{|c| c| c| c| c| c|}
\hline
 & $TTTT$ & $HHHH$ & $H'H'H'H'$ & $VVVV$ & $SSSS$   \\
\hline
$\mathcal{R}^4$ & $(E/f)^{12}$ & $(m/E)^4\,(E/f)^{12}$ & $(m/E)^4\,(E/f)^{12}$& $(E/f)^{12}$&$(m/E)^4\,(E/f)^{12}$  \\
$\Phi^4$ & $\lambda_L$ & $\lambda_L(E/m)^4$& $\lambda_L(E/m)^4$ & $\lambda_L(E/m)^8$ & $\lambda_L(E/m)^{12}$  \\
$\Phi^4$ tuned & $0$ & $0$&$\lambda_L(E/m)^4$ & $\lambda_L(E/m)^6$ & $\lambda_L(E/m)^{10}$  \\
\hline
\end{tabular}
\center
\caption{{\it Examples of leading energy-growth rate of $2\to 2$ scattering amplitudes with $E\sim\sqrt{s}\sim\sqrt{-t}$, for generic Lorentz contractions (the last line corresponding to the specific combination in Eq.~(\ref{tunedPotential})). For more general combinations of external polarizations, the energy-growth can be readily estimated from the examples in the table.}}
\label{tab:Scalings}
\end{table}

Interactions of the type $\mathcal{R}^4/f_T^{12}$ contribute like $g_T^2(E)\simeq (E/f_T)^{12}$ to the elastic scattering of \emph{transverse} polarizations.
Indeed, at the leading order in $m$, the Riemann tensor $\mathcal{R}$ sources only purely transverse polarizations (it is invariant in particular under the traceful gauge transformation that introduces the longitudinal fields in \eq{HcovDev2}).
The scattering of the $H, H'$ polarizations takes place only after mixing with the transverse ones, thus suppressed by powers of $m/E$, while for $V$ and $S$ the leading contributions arise from the first and second field redefinitions in (\ref{fieldRedef}), respectively.
Since $m/E<1$, transverse polarizations have the strongest coupling $g_T^2=\Lambda_T^{12}/f_T^{12}$ and lowest strong-coupling scale (associated to $\mathcal{R}^4$ only). The transverse cutoff is then bounded by
\begin{equation}\label{sct}
\Lambda_T\lesssim \Lambda^\text{sc}_T=(4\pi)^{1/6}f_T\,,
\end{equation}
which is very close to the scale $f_T$ which characterises the interactions and, a priori, arbitrarily far from the particle mass $m$. Therefore, for what concerns the transverse polarizations in a $\mathcal{R}^4$-theory, the ratio between mass and cutoff can take arbitrary values $0\leq\epsilon=m/\Lambda<1$ in the interacting theory.

\begin{figure}[t]
\begin{center}
\includegraphics[height=7cm]{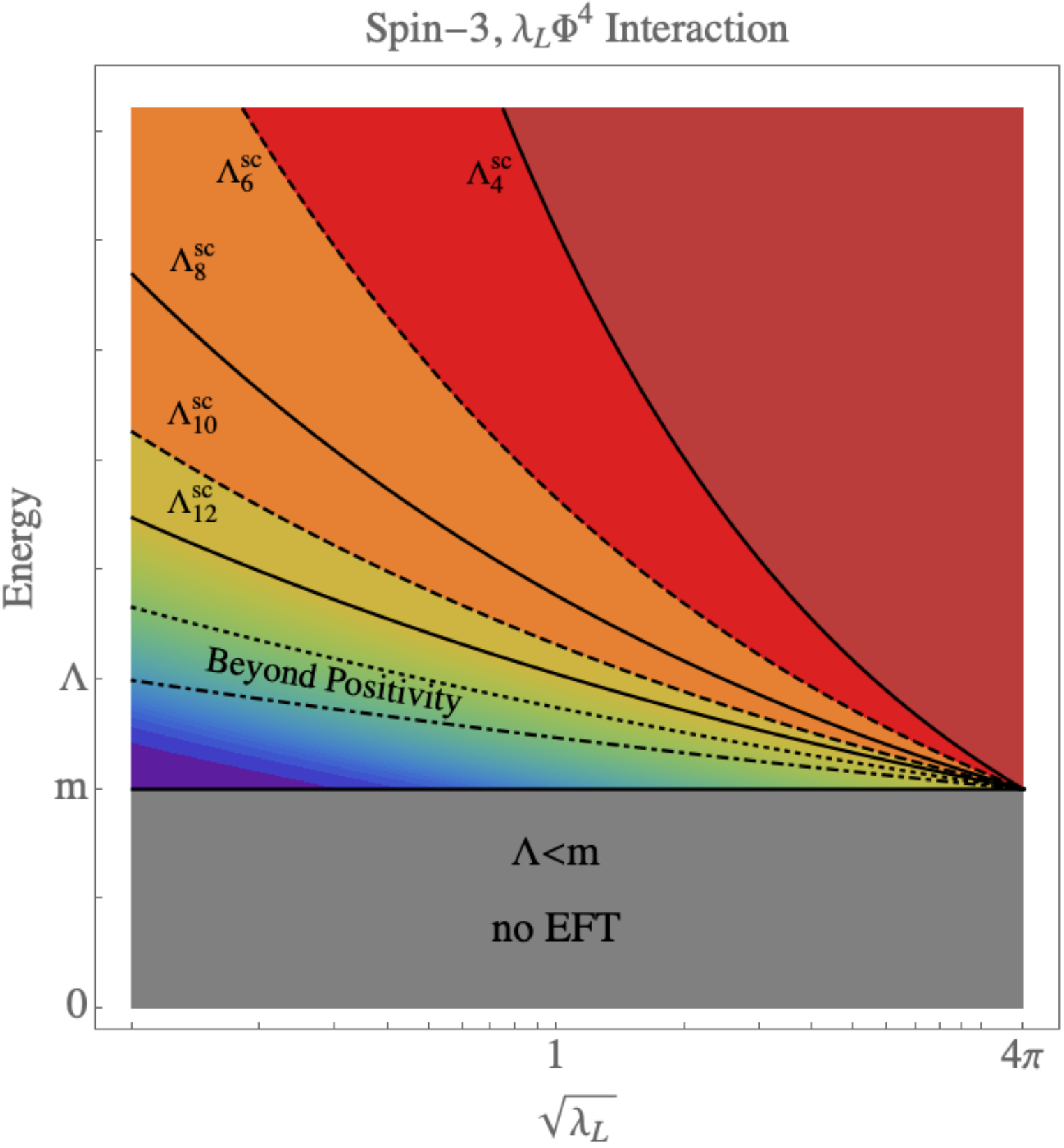}\hspace{1cm}
\includegraphics[height=7cm]{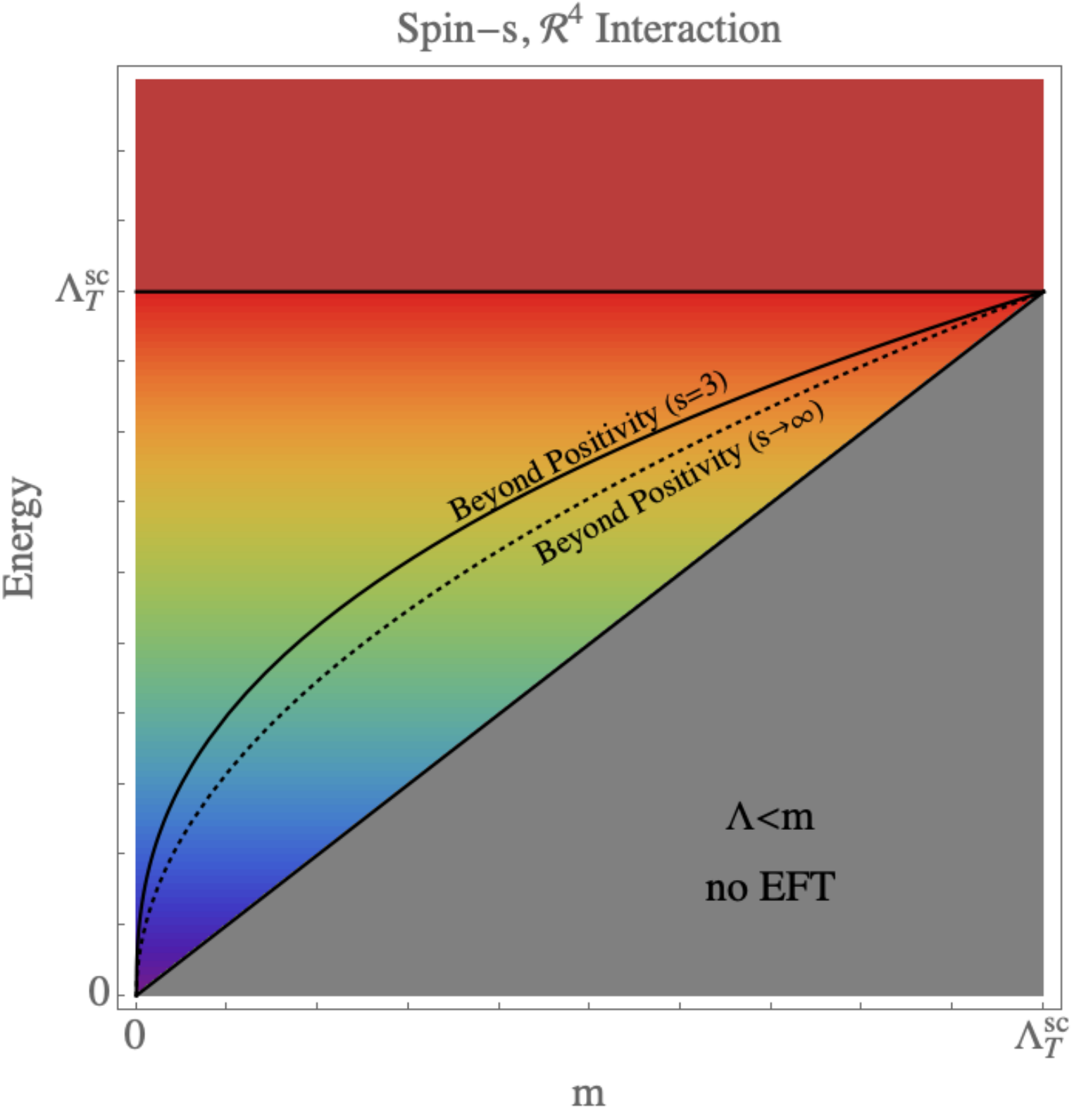}
\caption{{\it LEFT: Different strong-coupling scales  for $s=3$ as function of $\sqrt{\lambda_L}$ in \eq{Potential} or \eq{generalLagrangian}. For $E\gtrsim m$ the interaction strength increases at high energy; the $S(V)$ polarizations  are strongly coupled at  $\Lambda^\text{sc}_{12}(\Lambda^\text{sc}_{8})$ generically (solid lines) or at $\Lambda^\text{sc}_{10}(\Lambda^\text{sc}_{6})$ for the tuned theory (dashed lines). Beyond-positivity bounds  for the $\lambda_L\Phi^4$ interaction (dotted) and $\lambda_L\partial^4\Phi^4/\Lambda^4$ (dot-dashed): this represents the maximal cutoff $\Lambda$ of the theory, well below the strong-coupling scales. RIGHT: similar energy scales and beyond-positivity bounds for the $\mathcal{R}^4$ interaction as function of $m$ (solid line for $s$=3, dotted for $s\to\infty$). }}
\label{fig:SC}
\end{center}
\end{figure}

Interactions of the form $\lambda_L \Phi^4$ are instead very different. They source longitudinal polarizations that grow at high energy (see discussion below \eq{HcovDev2}) and generically become strongly coupled at energies different than $\Lambda_T^{sc}$.
For instance, the leading contribution from \eq{Potential} to the amplitude for scattering of the helicity-0 polarizations at $E\gg m$ is
\begin{equation}\label{fastestgrowth}
\mathcal{M}^{\small SS\rightarrow SS}\,= \frac{1}{25 m^{12}}\left[\frac{3}{4} \left(2\lambda_1-\lambda_2 + 2 \lambda_3 \right)(s t u)^2 + \frac{ \lambda_2 + 2\lambda_3}{16}\left(s^2+t^2+u^2\right)^3 \right] + \cdots
\end{equation}
and grows as fast as $\sim (E/m)^{12}$,  so that the theory becomes strongly coupled already at 
\begin{equation}
E\simeq m \left(\frac{16\pi^2}{\lambda_L}\right)^{1/12} = \Lambda^\text{sc}_{12}\,,
\end{equation}
where we defined a generic  strong-coupling scale
\begin{equation}
\label{strongScale}
\Lambda^\text{sc}_n\equiv m \left(\frac{16\pi^2}{\lambda_L}\right)^{1/n}\,.
\end{equation}
 Similarly, the $V$ polarizations become strongly coupled at $\Lambda^\text{sc}_8>\Lambda^\text{sc}_{12}$ while $H$-polarizations at  $\Lambda^\text{sc}_4>\Lambda^\text{sc}_8>\Lambda^\text{sc}_{12}$. Interactions involving different polarizations have other strong-coupling scales as a result of the different powers of mass in \eq{HcovDev2} and are illustrated in Figure~\ref{fig:SC}.

In light of this it is interesting to see that the choice $\lambda_2=\lambda_1=-2\lambda_3$ cancels the terms~$\sim E^{12}$ in \eq{fastestgrowth} and leads to amplitudes that grow only as~$\sim E^{10}$:
\begin{equation}
\label{TunedAmplitude}
\mathcal{M}^{SS\rightarrow SS} = \frac{4}{25} \frac{\lambda_3 }{m^{10}}s t u (s^2+t^2+u^2) + \cdots \,,
\end{equation}
with a similar cancellation  for helicity-1 amplitudes, see Table~\ref{tab:Scalings}. 
This tuning corresponds to the combination \cite{Bonifacio:2018vzv,Bonifacio:2017iry}
\begin{equation}
\label{tunedPotential}
-\lambda_L\Phi^4 = -\lambda_3\epsilon_{\mu_1\mu_2\mu_3\mu_4}\epsilon_{\nu_1\nu_2\nu_3\nu_4}\Phi^{\mu_1\nu_1\sigma}\Phi^{\mu_2\nu_2}_{\phantom{{\mu_2\nu_2}}\sigma}\Phi^{\mu_3\nu_3\rho}\Phi^{\mu_4\nu_4}_{\phantom{{\mu_4\nu_4}}\rho} \,,
\end{equation} 
and is analogous to what happens in the theory of massive gravity~\cite{ArkaniHamed:2002sp,deRham:2010kj} where it leads to the raising of the strong-coupling scale from $\Lambda_5$ to $\Lambda_3$.
The high-energy limit
\begin{equation}
m\rightarrow 0 \,,\qquad \lambda_L\rightarrow 0\,,\qquad \Lambda^\text{sc}_{10}= \text{fixed}\,,
\end{equation}
selects the interactions 
\begin{align}
\label{tunedInteractions}
\begin{split}
\frac{8 \lambda_3}{m^{10}} \epsilon^{\mu_1\mu_2\mu_3\mu_4}\epsilon^{\nu_1\nu_2\nu_3\nu_4}\Big[&\partial_\rho \hat{F}_{\mu_2\nu_2}\partial^\rho \hat{F}_{\nu_1\mu_1}\,\partial_{\mu_3\nu_3\sigma}\hat\pi\,\partial_{\mu_4\nu_4}^{\phantom{\mu_4\nu_4}\sigma}\hat\pi+2\partial_\sigma \hat{F}_{\mu_2\nu_2}\partial^\rho \hat{F}_{\nu_1\mu_1}\,\partial_{\mu_3\nu_3\rho}\hat\pi\,\partial_{\mu_4\nu_4}^{\phantom{\mu_4\nu_4}\sigma}\hat\pi\\
&-\frac{1}{25}\eta_{\mu_4\nu_4}\partial^\sigma\hat\pi \partial_{\mu_1\nu_1\sigma}\hat\pi\partial_{\mu_2\nu_2\rho}\hat\pi\partial_{\mu_3\nu_3}^{\phantom{{\mu_3\nu_3}}\rho}\hat\pi\Big]\,,
\end{split}
\end{align}
that are most relevant for $E<\lsc{10}$, where the tuned EFT is valid. 
Here only the $S$ and $V$ helicities are interacting, while the others decouple.

The tuning of the potential \eq{tunedPotential} can be equivalently seen in terms of the Goldstone field~$\pi$. At high energy, the interaction $\Phi^4$ corresponds to $\sim \left(\partial^3\pi\right)^4$, but  the specific combination appearing in Eq.~(\ref{tunedPotential}) vanishes up to total derivatives because of anti-symmetrization with the $\epsilon$-tensors. Therefore, the would-be sub-leading terms $\sim (\partial\pi)(\partial^3\pi)^3$ now dominate the scalar amplitude and reproduce the energy-growth $\sim E^{10}$ reported in Table \ref{tab:Scalings}. In fact, it is easy to show that each operator in Eq.~(\ref{tunedInteractions}) is a non-trivial invariant under the $N=2$ polynomial shift symmetry.\footnote{The four-field scalar operator in Eq.~(\ref{tunedInteractions}) is not present in the classification of Ref.~\cite{Hinterbichler:2014cwa}, as one can check comparing the energy-growth of the scattering amplitudes. See also Ref.~\cite{Griffin:2014bta} for an exhaustive non-relativistic classification.} Besides, they are not renormalized by loops, as can be understood by simple derivative counting, similarly to what happens for $N=1$ invariant Galileons \cite{Luty:2003vm,Goon:2016ihr}.\\

The arguments of this section can be swiftly generalized for arbitrary HS fields $\Phi_{\mu_1...\mu_s}$. Interactions of the type $\lambda_L\Phi^4$ lead to scalar scattering amplitudes growing as $\mathcal{M}\sim E^{4s}$ with a strong coupling scale $\Lambda^\text{sc}_{4s}$. However, softer behaviours can be achieved with the generalization of the tuning Eq.~(\ref{tunedPotential}) to spin-$s$ fields
\begin{align}
s \text{ even:} \,\qquad &\epsilon^{\mu_1...\mu_4}\epsilon^{\nu_1...\nu_4}\cdots\epsilon^{\rho_1...\rho_4}\, \Phi_{\mu_1\nu_1...\rho_1}\Phi_{\mu_2\nu_2...\rho_2}\Phi_{\mu_3\nu_3...\rho_3}\Phi_{\mu_4\nu_4...\rho_4}\,,\\
s \text{ odd:} \,\qquad &\epsilon^{\mu_1...\mu_4}\epsilon^{\nu_1...\nu_4}\cdots\epsilon^{\rho_1...\rho_4} \Phi_{\mu_1\nu_1...\rho_1\alpha}\Phi_{\mu_2\nu_2...\rho_2}^{\phantom{\mu_2\nu_2...\rho_2}\alpha}\Phi_{\mu_3\nu_3...\rho_3\beta}\Phi_{\mu_4\nu_4...\rho_4}^{\phantom{\mu_2\nu_2...\rho_2}\beta}\,.
\end{align}
These potentials lead to scalar amplitudes $\mathcal{M}\sim E^{3s}$ or $\mathcal{M}\sim E^{3s+1}$ for even and odd spins respectively, realizing explicitly the optimal high-energy behaviour conjectured in Ref.~\cite{Bonifacio:2018vzv}. This result can be understood by consistently taking the decoupling limit, as explicitly shown in Appendix~\ref{appendix:tunedPotential}.
For the EFT to be perturbative, the cutoff $\Lambda_L$ must lie below the strong-coupling scale
\begin{align}
\textrm{Generic:} \,\qquad &\Lambda_L\lesssim\lsc{4s} \,, \qquad\qquad \label{gensgen} \\
\textrm{Tuned ($s$ even or odd):} \,\qquad &\Lambda_L\lesssim\lsc{3s\,,\,3s+1}\,. \qquad\qquad \label{genstun}
\end{align}
The vector polarizations are strongly coupled at $\Lambda^\text{sc}_{4(s-1)}$ and similarly for the other modes.

Interactions $\mathcal{R}^4/f_T^{4s}$, with $\mathcal{R}$ the Riemann tensor for spin-$s$,  imply amplitudes involving transverse polarizations that also grow as $\mathcal{M}\sim E^{4s}$, but are not suppressed by inverse powers of the mass. The strong-coupling scale of the transverse interactions is therefore mass independent,
\begin{equation}\label{scts}
\Lambda_T\lesssim \Lambda^\text{sc}_T=(4\pi)^{\frac{1}{2s}} f_T\,.
\end{equation} 
Other polarisations have larger cutoffs associated with the $\mathcal{R}^4/f_T^{4s}$ interactions.

\subsection{Positivity}
\label{sec:dispRelations}

For a given mass $m$, the longitudinal scalar modes remain perturbative only up to energies of order $\lsc{}$. 
If the underlying  UV completion is Lorentz invariant, unitary, casual and local, one can obtain stronger bounds on the physical cutoff $\Lambda$~\cite{Adams:2006sv,Bellazzini:2016xrt,Nicolis:2009qm}, which may be pushed well below $\Lambda^\text{sc}$, as we will now show.
The key physical quantity that enters these arguments is the elastic $2\to2$ scattering amplitude $\mathcal{M}^{z_1z_2 z_1 z _2}(s,t)$,  with (linear) polarizations labelled by  $z_i$, which in the forward elastic limit $t=0$
enters into the $n$-subtracted ``IR residue'', namely\footnote{In this equation, the contour of integration $\Gamma$ encloses all the physical IR poles $s_i$ associated with stable resonances, if any,  together with the point $\mu^2< 4m^2$. See Refs.~\cite{Adams:2006sv,Bellazzini:2017fep} for more details. }
\begin{equation}
\label{Sigmadef}
\Sigma_{\mathrm{IR}}^{z_1 z_2\,(n)} \equiv \frac{1}{2\pi i} \oint_\Gamma ds\frac{\mathcal{M}^{z_1z_2}(s)}{(s-\mu^2)^{n+1}} =\sum \underset{s=s_i,\mu^2}{\mathrm{Res}}\left[ \frac{\mathcal{M}^{
z_1 z_2}(s)}{(s-\mu^2)^{n+1}} \right] \,,
\end{equation}
with $\mathcal{M}^{z_1 z_2}(s)\equiv \mathcal{M}^{z_1 z_2 z_1 z_2}(s,t=0)$. The $\Sigma_{\mathrm{IR}}^{z_1 z_2\,(n)}$ is calculable within the EFT, since $0<\mu^2 < 4m^2$, in terms of the couplings and masses of the IR theory. 
 Using the analytic properties of the scattering amplitude, the Froissart-Martin asymptotic bound, crossing symmetry and the optical theorem, one derives the dispersion relation for even $n\geq 2$
\begin{align}
\label{eq:dispersive}
\Sigma_{\mathrm{IR}}^{z_1 z_2\,(n)} = & \sum_X
\int^{\infty}_{4m^2} 
\frac{ds}{\pi} \sqrt{1-4\frac{m^2}{s}} \left[\frac{s \sigma^{z_1 z_2\rightarrow X}(s) }{(s-\mu^2)^{n+1}}   +\frac{s \sigma^{-\bar{z}_1 z_2 \rightarrow X }(s) }{(s-4m^2+\mu^2)^{n+1}}\right] \,,
\end{align}
which connects the IR physics (matched, by definition, with the EFT) to the UV, through an integral of the total cross section for the production of any  (not necessarily elastic) kinematically accessible state $X$. In any interacting theory the right-hand side of Eq.~(\ref{eq:dispersive}) is strictly positive,
\begin{equation}\label{positivityresiduen}
\Sigma_{\mathrm{IR}}^{z_1 z_2\,(n)}(\mu^2) >0\,,
\end{equation}
for any value of $\mu^2$ in the above range. 
For instance,  the $TTTT$ amplitudes  from a generic 
$\mathcal{R}^4/f_T^{4s}$ interaction scale like $\mathcal{M}\sim s^{2s}$ so that  the residue
\begin{equation}\label{IRresiduespins}
\Sigma^{TT  (n)}_{\mathrm{IR}}\sim \frac{m^{4s -2n}}{f^{4s}_T} 
\end{equation}
is proportional to some power of $m$.  
In Section~\ref{sec:ConstraintSpin3} we show that these positivity bounds set very strong constraints on the EFT of HS,  even stronger than in the case of massive gravity~\cite{Bellazzini:2017fep,deRham:2018qqo,Cheung:2016yqr}. 

\paragraph{Beyond Positivity.}
The integral on the right-hand side of \eq{eq:dispersive} contains a positive IR contribution for $4m^2 < s \lesssim \Lambda^2$ that is still calculable within the EFT.\footnote{\label{ftnt}Strictly speaking, at $E\sim\Lambda$ the EFT produces results which are $O(1)$ accurate in the dispersion relation. Better accuracy can be derived by using the EFT only up to $E^{max}< \Lambda$, as in Refs.~\cite{Bellazzini:2017fep,Distler:2006if,deRham:2017imi}. While it is straightforward to keep track of this factor, it is not very important since even $O(1)$ errors in the dispersion relations translate into small modifications of the bounds on the cutoff for large spin. } The unknown UV contribution $s\gg \Lambda^2$ is still positive and Eq.~(\ref{eq:dispersive}) can be turned into an inequality
\begin{equation}
\label{eq:bound}
 \Sigma_{\mathrm{IR}}^{z_1 z_2}  \!>\!  \sum_X
\int^{\Lambda^2} 
\!\!\!\frac{ds}{\pi s^2} \left[ \sigma^{z_1 z_2\rightarrow X}(s)  +\sigma^{z_1 -\bar{z}_2 \rightarrow X }(s) \right]_{\mathrm{EFT}}\,,
\end{equation}
where we focus  on the $n=2$ residue for $\mu^2\sim m^2\ll\Lambda$.

Consider first  spin-$s$ interactions of the type $\mathcal{R}^4/f_T^{4s}$. The elastic cross section for $TTTT$ scattering scales as $\sigma \sim 1/16\pi^2\times s^{4s-1}/f_T^{8s}$ at high energy, while the $n=2$ residue \eq{IRresiduespins} is suppressed by $4s-4$ powers of the mass.
Using these ingredients in \eq{eq:bound} we find parametrically
\begin{equation}\label{ttttbound}
 \Lambda\lesssim  \left(16 \pi^2f_T^{4s}m^{4s-4}\right)^{\frac{1}{8s-4}}= \Lambda_T^{sc}  \left(\frac{m}{\Lambda_T^{sc}}\right)^{\frac{4s-4}{8s-4}}< \Lambda_T^{sc}\,.
\end{equation}
This new cutoff scale 
is always smaller than the strong-coupling scale \eq{scts}, unless the EFT is not valid $m\sim \Lambda_T^{sc}$, as illustrated in the right panel of Fig.~\ref{fig:SC}.

For interactions of the type $\lambda_L\Phi^4$, the strongest constraints come from studying the longitudinal amplitudes $SSSS$, whose forward limit generically scales as $\lambda_L s^{2s}/m^{4s}$. The bound \eq{eq:bound} then implies
\begin{equation}\label{llllbound}
 \Lambda\lesssim   m \left(\frac{16 \pi^2}{\lambda_L}\right)^{\frac{1}{8s-4}}=\Lambda_{8s-4}^{sc}< \Lambda_{4s}^{sc}
\end{equation}
which is, again, lower than the strong-coupling scales mentioned in \eq{gensgen}. For spin-3, these beyond-positivity bounds are illustrated  in Fig.~\ref{fig:SC}.

For tuned interactions the arguments are very similar, and beyond positivity leads to a cutoff $\Lambda<\lsc{6s-4}$ or $\Lambda<\lsc{6s-2}$ for even or odd spin respectively, still lower than the estimates in \eq{genstun}. We will see in Sec.~\ref{sec:ConstraintSpin3} that a radical impeachment will imply even stronger bounds, see Eq.~(\ref{BPBspin3}).

\paragraph{Beyond Positivity and Weak Coupling.} 
The dispersion relation \eq{eq:dispersive} can be used to relate residues with different numbers of subtractions. Neglecting for simplicity $\mu^2\sim m^2\ll\Lambda^2$ and working with linear polarizations, we define a subtracted residue
\begin{equation}\label{SigmaSub}
\widetilde{\Sigma}_{\Lambda^2}^{(n)} \equiv \Sigma_{\mathrm{IR}}^{(n)}-\frac{2}{\pi}\int_{4m^2}^{\Lambda^2} \frac{ds}{s^n} \sigma=\frac{2}{\pi}\int_{\Lambda^2}^{\infty} \frac{ds}{s^n} \sigma
\end{equation}
which by the Cauchy theorem is nothing but the anti-clockward integral over two half-circles,\footnote{For illustration, considering $n=2$ and $\mathcal{M}(s,t=0)=a(\mu^2)s^2+\frac{\beta}{2}s^2\left[\log(s/\mu^2)+ \log(-s/\mu^2)\right]$, then $\widetilde{\Sigma}_{\Lambda^2}^{(2)}= a(\mu^2)+\beta \log(\Lambda^2/\mu^2)$, which represents the run $s^2$-coefficient at the scale $\Lambda^2$. Incidentally, the positivity of the total cross section $\sigma$ shows that running from $\Lambda$ to $\Lambda^{\prime}< \Lambda$ makes the Wilson coefficient larger, that is $\widetilde{\Sigma}_{\Lambda^{\prime\,2}}^{(n)}>  \widetilde{\Sigma}_{\Lambda^2}^{(n)}$ for $\Lambda^\prime<\Lambda$, or equivalently $\beta = d a(\mu^2)/d\log\mu^2 <0$.} just above and below the branch cuts, centered at $s=0$ and of radius $\Lambda^2$. 
Since $s>\Lambda^2$ inside the integral in Eq.~(\ref{SigmaSub}), we have that
\begin{equation}\label{diffn}
\widetilde{\Sigma}_{\Lambda^2}^{(n)}>  \Lambda^4 \widetilde{\Sigma}_{\Lambda^2}^{(n+2)}\,.
\end{equation}
Now,  Eqs.~(\ref{ttttbound},\ref{llllbound}) imply that a sizeable separation between the mass $m$ and the cutoff $\Lambda$ is possible only if the theory is \emph{weakly} coupled~$\lambda_L,g_T\ll 4\pi$. Therefore, we can  calculate  $\widetilde{\Sigma}_{E^2\lesssim \Lambda^2}^{(n)}$ using the IR EFT, with (\ref{diffn}) setting  non-trivial bounds on the EFT coefficients. 
Just to make this apparent and direct, let us make the simplification of dropping the difference between $\widetilde{\Sigma}_{\Lambda^2}^{(n)}$ and $\widetilde{\Sigma}_{4m^2}^{(n)}=\Sigma_{\mathrm{IR}}^{(n)}$, which is neglecting the IR branch cuts in the dispersive integral relative to the UV ones.\footnote{See also Ref.~\cite{futureworkER} for a derivation that does not rely on ignoring the IR part of the integral, and Ref.~\cite{Englert:2019zmt} for a discussion in the context of the 2-point functions.}
This provides an extremely powerful constraint for the soft amplitudes typical of HS theories, which have the first few even powers in $s$ suppressed. Consider the example of a single operators of the form $\lambda_L\Phi^4$ which gives  $\widetilde{\Sigma}_{E^2\lesssim \Lambda^2}^{(n\leq 2s)}\simeq \lambda_L/m^{2n}$, or a single $\mathcal{R}^4/f_T^{4s}$ that gives $\widetilde{\Sigma}_{E^2\lesssim \Lambda^2}^{(n\leq 2s)}\sim {m^{4s -2n}}/{f_T^{4s}}$. Then, for  $n\leq 2s-2$ \eq{diffn} reads
\begin{align}\label{killed}
\lambda_L\Phi^4\!: &\quad  \lambda_L\frac{1}{m^{2n}}\gtrsim \lambda_L \frac{ \Lambda^4}{m^{2n+4}}\\
\mathcal{R}^4/f_T^{4s}\!: &\quad  \frac{m^{4s -2n}}{f_T^{4s}}\gtrsim \frac{ \Lambda^4 m^{4s -2n}}{m^4f_T^{4s}} 
\end{align}
so that in either case 
\begin{equation}
\label{killer}
m \gtrsim \Lambda
\end{equation}
in contradiction with the very assumption that the EFT has a well-defined range of validity! 
More generally, we expect Eq.~(\ref{killed}) and such a conclusion to hold true even when the IR contribution in $\widetilde{\Sigma}_{\Lambda^2}^{(n)}$ is retained, except that $\lambda_L$ is evaluated at $\Lambda$ rather than at $m$, as discussed in footnote 14.

\subsection{Constraints on spin-3}\label{sec:ConstraintSpin3}

\begin{figure}[t]
\centering
\includegraphics[width=1\textwidth]{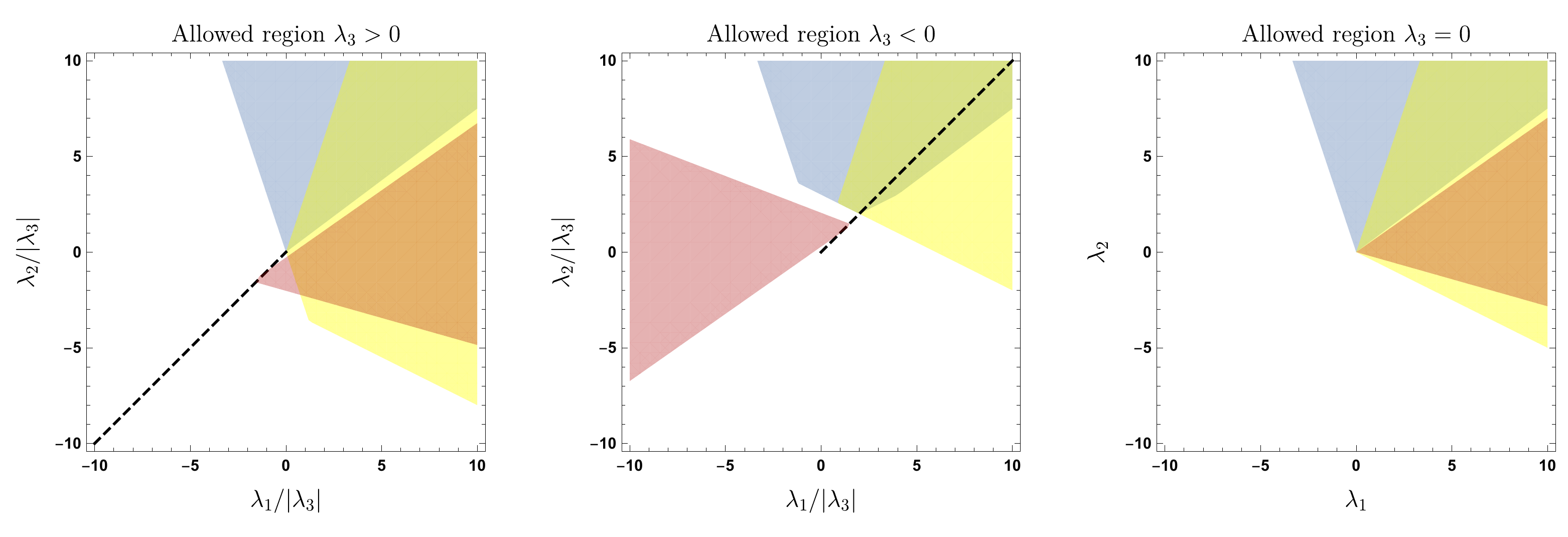}
\caption{{\it Allowed regions of the coefficients $\lambda_{1},\lambda_{2}$ as function of the sign of $\lambda_3$. Blue: result from scattering of linear-definite polarizations \eq{pos1}. Red: result from $\Sigma^{(2)}_\text{IR}>0$ by scattering different choices of linear combination of polarizations. Yellow: result from $\Sigma^{(4)}_\text{IR}>0$ by scattering only linear-definite polarizations. The dotted line correspond to the tuning $\lambda_1 =\lambda_2= -\lambda_3$ in \eq{tunedPotential}.}  }\label{fig:positivitesPotential}
\end{figure}

As an example of the general arguments given above, we focus here on the spin-3 case and the interactions in \eq{Potential}. The non-vanishing elastic residues for $n=2$ at the crossing-symmetric point $\mu^2=2m^2$ read
\begin{align}
\Sigma_{IR}^{VV}&=\frac{16}{75m^4}(-2 \lambda_1+7\lambda_2+6\lambda_3)>0\,, & \Sigma_{IR}^{VS}&=-\frac{16}{75m^4}(3\lambda_1-4\lambda_2)>0\nn\\
\Sigma_{IR}^{VV'}&=-\frac{32}{225m^4}(3\lambda_1-5\lambda_2)>0\,,&\Sigma_{IR}^{VH} &= \frac{8}{{15 m^4}}\lambda_2>0\label{pos1}\\
\Sigma_{IR}^{SS}& = -\frac{18}{25m^4}(\lambda_1-2\lambda_2-2\lambda_3)>0\,,& \Sigma_{IR}^{SH}& = -\frac{2}{15m^4}(3\lambda_1-7\lambda_2)>0\nn\\
\Sigma_{IR}^{HH}& = \frac{1}{3m^4}(\lambda_1+2\lambda_2+6\lambda_3)>0\nn
\end{align}
while  $TX \rightarrow TX$ gives $\sim O(s)$ forward amplitudes for any state $X$ ($O(s^0)$ for $X=T$) and have vanishing residue.
The positivity constraints for $S,V,H$ helicities \eq{positivityresiduen}  selects the  blue (yellow) regions for $n=2(4)$ in Fig.~\ref{fig:positivitesPotential}.
Moreover, generic linear combinations of polarizations $X^1X^2\rightarrow X^1X^2$ with
$X^{1,2}_{\mu\nu\rho} = \sum_{i=1}^7 x^{1,2}_i \epsilon^i_{\mu\nu\rho}$ and  $\sum_{i=1}^7 x^{1,2}_i x^{1,2}_i = 1$, lead to more constraints (indeed $\Sigma_{IR}^{XY}$ is not just a linear combination of \eq{pos1} but it includes inelastic residues summed into an elastic combination). 
%
Such constraints are linear in $\vec{\lambda}=(\lambda_1,\lambda_2,\lambda_3)$ and can be written as
$\Sigma_{IR}^{XY}\equiv \vec{F}(\mu^2,\vec{x^1},\vec{x^2})\cdot\vec{\lambda} >0$ for a certain function $\vec{F}$, that has to hold for any  $\mu^2 \in [0,4m^2]$ and polarizations $\vec{x^{1,2}}$. With 12 free variables $x_i^{1,2}$ and only 3 coefficients $\lambda_i$, we find numerically a finite set of points $\{\mu^2_n,\vec{x}_n,\vec{y}_n\}$ implying the positivity constraints shown in red in Fig.~(\ref{fig:positivitesPotential}). The lack of  overlap between the red and blue regions implies that the theory with an infrared dominating potential is inconsistent.\footnote{See Ref.~\cite{Bonifacio:2016wcb} for similar bounds on a restricted class of massive spin-2 theories.}

Surprisingly, the leading interactions \eq{Potential}  are incompatible with positivity when all helicity amplitudes are taken into account. This does not mean that the entire HS formulation is inconsistent, but implies that the leading consistent interactions are not those of \eq{Potential}, but must be more irrelevant, i.e.~higher in derivatives.\footnote{The interactions of \eq{Potential} can still be present, but they must be subdominant or at most comparable in the IR; since they are the most relevant, they are also subdominant at higher energy.  Keeping them does not change our qualitative conclusions.}
As a matter of fact however, more irrelevant interactions $p \geqslant 2$ (even)
\begin{equation}
\label{generalLagrangian}
\mathcal{L} = \sum_p  \lambda_L^{(p)}\, \frac{\partial^{p}}{\Lambda^{p}}\Phi^{4}  = \lambda_L^{(0)}\Phi^4+\frac{\ \lambda_L^{(2)}}{\Lambda^2} \partial^2\Phi^4+\frac{ \lambda_L^{(4)}}{\Lambda^4} \partial^4\Phi^4 + \cdots
\end{equation}
 lead to even stronger bounds. Up to $O(1)$ factors, \eq{eq:bound} gives 
\begin{align}
\label{BPBspin3}
\textrm{Generic:} \,\qquad &\Lambda \lesssim \lsc{8s-4+p}\,, \qquad\qquad \\
\textrm{Tuned ($s$ even or odd):} \,\qquad &\Lambda \lesssim \lsc{6s-4+p\,,\,6s-2+p}\,, \qquad\qquad
\end{align}
always stronger than the analog bound in the theory without derivatives \eq{llllbound}: $\lsc{8s-4+p}<\lsc{8s-4}$ and analogously for the tuned case.

We have studied numerically the 24-dimensional parameter space of operators of the form $\partial^2\Phi^4$ looking for combinations that satisfy the simple positivity bound $\Sigma^{(n)}>0$ for $n=2,4,6$ for all elastic amplitudes. We find that no linear combination passes all positivity requirements. 
This suggests that the most important unitarity-consistent self-interaction at low energy is actually much more irrelevant than naively anticipated, hence leading to even more stringent beyond-positivity bounds.
For illustration we show the bound \eq{BPBspin3} for operators $p=4$ as a dot-dashed line in the left panel of Fig.~\ref{fig:SC}. 
With similar tools, we have analysed few of the many contractions $\mathcal{R}^4$, without  finding any combination that passes all the simple positivity bounds. This is certainly very intriguing and we leave for future studies a systematic discussion of more irrelevant operators.

\subsection{Couplings to other fields}\label{sec:matter}

So far we have focussed on HS self-interactions. From a phenomenological perspective, interactions with other fields are very important as well. These are interesting in the context of cosmology~\cite{Bordin:2018pca,Arkani-Hamed:2018kmz}, but also in collider physics. For instance, the highly irrelevant HS interactions would explain the absence of BSM signals at low energies, compatibly with the presence of detectable structure at higher energies \cite{Liu:2016idz,Bellazzini:2017bkb,Bellazzini:2018paj}.
For the same reason, one can wonder whether HS may provide viable dark matter candidates~\cite{Bruggisser:2016ixa,Bruggisser:2016nzw}, where e.g.~the WIMP miracle is realized because of the irrelevance of the interactions at low energies rather than a genuine weak coupling.

\paragraph{Linear Couplings.}
We consider in the following the case where the sector that gives rise to the HS longitudinal modes contains as well some of the other matter fields, e.g.~fields of the SM (fermions $\psi$'s, gauge bosons $A$, etc.). These enter quadratically in the dimension-$5$ current for the spin-3, which reads\footnote{The universal gravitational coupling to the energy-momentum tensor is present as well, but this can be neglected as long as $\epsilon = m / \Lambda$ and the other couplings are not too small, e.g.~for $\lambda_L \gg \Lambda^2/m^2_{\mathrm{Pl}}$. Moreover, it can be consistently subtracted by the positivity bounds~\cite{Bellazzini:2019xts}. }
\begin{equation}
\label{mattercurrentcoup}
\frac{\Phi_{\mu\nu\rho}}{\Lambda^2}\mathcal{J}_{\mu\nu\rho}\big|_\mathrm{mat}=\frac{\Phi_{\mu\nu\rho}}{\Lambda^2}\sum_{\mathrm{sym}} \left\{c_\psi\left(\bar{\psi}\gamma_\nu\overset{\leftrightarrow}{\partial_{\nu}}\overset{\leftrightarrow}{\partial_\rho} \psi-\frac{1}{5}(\partial_\mu\partial_\nu -\square \eta_{\mu\nu})\bar{\psi}\gamma_\rho \psi \right) + c_A F^{+}_{\nu\alpha}\overset{\leftrightarrow}{\partial_{\mu}}F^{-}_{\alpha\rho}+\cdots \right\}\,,
\end{equation}
where the sum $\sum_{\mathrm{sym}}$ is over all permutations (normalized by their number),  and $F_{\mu\nu}=F^{+}_{\mu\nu}+F^{-}_{\mu\nu}$, $1/2\epsilon_{\mu\nu\rho\sigma}F_{\rho\sigma}=F^{+}_{\mu\nu}-F^{-}_{\mu\nu}$ \cite{Anselmi:1998bh}.  These matter spin-3 currents are conserved in the free theory whereas they are not in the interacting theory, in agreement with  the Coleman-Mandula theorem. The non-conservation is  a non-issue, as long as the cutoff of the longitudinal modes is not lowered.
For example, the scattering of a pair of HS particles into matter that follows from \eq{mattercurrentcoup} scales as
\begin{equation}\label{eq:inelstaicmatter}
\mathcal{M}^{SS\rightarrow \bar{\psi}\psi, AA,\ldots}\sim c^2_{\psi,A,\dots} \left(\frac{E^{10}}{m^6\Lambda^4} \right) \,,
\end{equation}
which exhibits  the same energy-growth as the tuned $\Phi^4$-potential \eq{tunedPotential} (see Table~\ref{tab:Scalings}). Therefore, the strong-coupling scale associated with \eq{mattercurrentcoup} is unchanged as long as $c^2_{\psi,A,\dots} < \lambda_L\Lambda^4/m^4$. 
On the other hand, the beyond-positivity bounds \eq{eq:bound}  become stronger by retaining the inelastic channel \eq{eq:inelstaicmatter}  on the right-hand side of the dispersion relation for the $SSSS$ amplitude, and imply the scaling 
\begin{equation}
c_{\psi,A,\dots} \lesssim O(\lambda^{1/4}_L m^2)\,.
\end{equation}
For $\lambda_L\sim m^4$, this gives $c_{\psi,A,\dots} \sim O(m^3)$ and therefore the coupling of the longitudinal zero mode $\pi$ to matter would be finite in the massless limit.  
However, since the beyond-positivity bounds imply that $\lambda_L$ goes to zero even faster than $m^4$, $\pi$ should actually decouple from matter in the massless limit. 

\paragraph{Quadratic Couplings.} 
Other HS-matter couplings can be built e.g.~with $\Phi^2$ or $\mathcal{R}^2$ and singlet operators from matter fields; these are the dominant interactions preserving the  $\Phi_{\mu\nu\rho}\rightarrow - \Phi_{\mu\nu\rho}$ symmetry necessary for the HS to possibly play the role of dark matter.
 In this class, the coupling to a scalar (e.g.~the Higgs boson) admits even marginal couplings 
\begin{equation}\label{marginalinteractions}
g_L^2\frac{m^2 |\mathcal{H}|^2}{\Lambda^2}\left( c_{1,H}\Phi_{\mu\nu\rho}^2-3 c_{2, H}\Phi_\mu^2\right)\,,
\end{equation}
where we have inserted  $m^2$ as discussed below \eq{generalCovD}.
If $\mathcal{H}$ obtains a vacuum expectation value $\langle v\rangle$, as the Higgs in the SM, this interaction contributes to the HS potential, detuning it, and could  lower the cutoff of the HS theory to $\Lambda^2_{\mathrm{ghost}}=\Lambda m^2/(v g_L \sqrt{|c_{1, H}-c_{2,H}|})$. 
This contribution can be removed by tuning $c_{1, H}=c_{2,H}$.  Alternatively, using again the non-elastic channel $SS\rightarrow \mathcal{H}\mathcal{H}$ on the right-hand side of the dispersion relation \eq{eq:bound}, we see that for generic $c_{i, H}$
\begin{equation}
\label{scalinggL}
g_L\lesssim O(\lambda_L^{1/4} m) \,,
\end{equation} 
which implies that, even in the presence of \eq{marginalinteractions}, the cutoff is   expected to be $\Lambda\ll \Lambda_{\mathrm{ghost}}$, given that $\lambda_L^{1/4}$ happens to scale faster than $m$. 

Other invariants, of higher dimensionality but contributing to interactions with different helicity structure, can be built with $\mathcal{R}^2$ or other SM operators, for instance
\begin{equation}\label{opsDM}
g_T^2\frac{|\mathcal{H}|^2 \mathcal{R}^2}{\Lambda_T^{2s}}\,,\quad g_T^2 \frac{F_{\mu\nu}^2 \mathcal{R}^2}{\Lambda_T^{2s+2}}\,, \quad g_L^2 \frac{m^2F_{\mu\nu}^2 \Phi^2}{\Lambda_L^{8}}\,.
\end{equation}

%
%

\section{Conclusions and Outlook}

In this article, we have provided an effective quantum field theory description of abelian, single flavor, self-interacting massive (integer) higher-spin states. 
The relativistic degrees of freedom of the HS correspond to the longitudinal (Goldstone) and transverse (gauge) modes, which  follow different power counting rules since they realize, non-linearly, different symmetries. The separation into longitudinal and transverse modes is both conceptual and practical. It offers, for example, a neat understanding of the structure of the HS kinetic and mass terms needed to generate a gap between the mass of the HS and the cutoff of the theory, by removing would-be light ghosts from the spectrum. Moreover, the symmetries of the modes have allowed us to identify the least irrelevant interactions that come in a variety of structures, depending on the helicities involved.  For example, the leading operators that contribute to scattering amplitudes among transverse-only modes are  made of the HS-Riemann tensor $\mathcal{R}^n$ (which respects the emerging gauge symmetry of the massless limit),  whereas the  leading one for scattering Goldstone-only modes is of the form $\Phi^n$. 
In between, there are other operators, e.g.~$\mathcal{R}^2 \Phi^2$, which dominate mixed helicity scatterings. These represent the HS generalization of the massive spin-1 $F_{\mu\nu}^4$, $A_\mu^4$, and $F_{\mu\nu}^2 A^2_\rho$ type of operators, respectively. 

This EFT may be useful for phenomenological applications. Indeed, heavy HS can have interesting signatures in cosmology, through their imprint  in the cosmic microwave background. Lighter HS coupled to the SM fields could in principle be observed at colliders or could play the role of dark matter: their irrelevant interactions are very small at low energies, potentially explaining why they have not yet been observed.

As for any relativistic EFT to make sense, a gap between the mass of the HS states and the cutoff of the theory is necessary. 
We have studied whether such a gap may originate from underlying microscopic UV completions that are causal, local and unitary, a question that can be addressed by using dispersion relations. 
Our findings are summarised in Fig.~\ref{fig:SC} and show that, for a given strength of the 4-point interaction, the cutoff is parametrically close to the mass, and it goes to zero in the limit where the HS states are massless. 
As representative of the general method, we find an upper limit on the cutoff of a theory with $\mathcal{R}^4/f_T^4$ interactions by studying $TTTT$ amplitudes, see  \eq{ttttbound}.  
For interactions among the longitudinal modes controlled by $\lambda_L\Phi^4$, we have studied the $SSSS$ scattering amplitudes, leading to the upper bound \eq{llllbound}.  
These bounds are always more stringent than those associated with the strong-coupling scales of the theory, Eqs.~(\ref{gensgen}-\ref{scts}), as portrayed by the lowest lying curves in Fig.~\ref{fig:SC}.
Alternatively, for a given cutoff, the interaction strength must vanish sufficiently fast as the mass goes to zero, and the theory quickly becomes free. 

An even stronger bound, \eq{killer}, can be obtained whenever the IR theory is more weakly coupled than the UV completion: it requires the mass of the HS to be as large as the cutoff, invalidating the EFT, under our assumptions.
 
These arguments hold for general spin but rely on estimates based on dimensional analysis. To make the bounds more concrete and precise, we have worked out the details of the explicit spin-3 case. By studying the most relevant interactions of the form $\lambda_L\Phi^4$, we have found a special combination that maximises the strong-coupling scale \eq{tunedPotential}, in agreement with the conjecture of Ref.~\cite{Bonifacio:2018vzv}. 
Surprisingly, however, both tuned and generic interactions of the form $\lambda_L\Phi^4$ do not pass standard positivity constraints that use mixed-helicity elastic amplitudes. This means that spin-3 self-interactions are actually more irrelevant than one would have naively anticipated, given that the would-be leading ones are very much suppressed. Neither the 24-dimensional space spanned by the couplings of  irrelevant operators of the type $\lambda_L\partial^2\Phi^4/\Lambda^2$ is consistent  with our positivity bounds.  
We find intriguing the lack of any consistent interaction at the order we have studied, perhaps a sign of a deeper inconsistency, the study of which we leave for future work.


%

Our bounds are general and robust because they are derived from fundamental properties of the S-matrix together  with basic EFT reasoning.  However, one can try to relax the assumptions that go into the EFT. 
Perhaps the most obvious direction would be  to take $m\simeq \Lambda$ at face value by adding extra states of lower spin at around the mass of the highest spin state, trying to construct a new EFT for this larger set of degrees of freedom. The structural question would then become whether a finite set of degrees of freedom is needed to generate a new gap $m/\Lambda\ll 1$ consistent with the positivity bounds. 
For instance, an odd spin  $\Phi_s$ may couple to a lower spin $\Phi_{s-1}$ in order to form trilinears $\frac{m_s^2 m_{s-1}}{\Lambda^2} \Phi^2_{s}\Phi_{s-1}$, $m^3_{s-1} \Phi^3_{s-1}$, etc.~that affect significantly the positivity bounds for $m_s\simeq m_{s-1}$. Alternatively, one can relax the discrete symmetry $\Phi_s\rightarrow -\Phi_s$ for even spin and consider cubic vertices $\frac{m^3}{\Lambda^2} \Phi_s^3$, which give $\mathcal{M}^{SS} \sim (m/\Lambda)^4 \left(s /m^2\right)^{3s-1}$ in the hard scattering limit, whereas providing $O(1)$ effects to the IR residues, relative to the contribution from $\Phi^4$, in the positivity bounds.  However, while trilinear couplings may resolve the inconsistency of the $\Phi^4$ interaction with the standard positivity bounds,  more stringent constraints than \eq{llllbound} are expected by the beyond-positivity bounds. A detailed analysis of these alternative EFTs is left to future work.

\subsection*{Acknowledgments}
We would like to thank James Bonifacio for useful discussions. F.S.~thanks IPhT for the hospitality during the very early stages of this work.
F.R.~is supported by the Swiss National Science Foundation under grant no.~PP00P2-170578. J.S.~has been supported by the DFG Cluster of Excellence 2094 ``Origins'' and by the Collaborative Research Center SFB1258.

\begin{appendices}

\numberwithin{equation}{section}

\section{Tuning conditions}
\label{appendix:tuning}

In this appendix, we show that the mixing between the spin-$s$ field and the current of Goldstones $\mathcal{J}$ generates the correct kinetic term for $\phi_{s-2}$ and $\phi_{s-3}$ if the mass terms and kinetic mixings of the auxiliary fields are tuned to specific values. These correspond to the coefficients found in Ref.~\cite{Singh:1974qz} after some field redefinitions.

Consider the resulting Lagrangian of a massive spin-$s$ particle Eq.~(\ref{totLagrangianHS}) after the transformation Eq.~(\ref{fieldRedefSpinS})
\begin{align}
\begin{split}
\mathcal{L} &=
\Phi \cdot \hat{\Gamma}_s +\phi_{(s-1)}\cdot \hat{\Gamma}_{s-1}+\frac{m^2}{2}(2s-1) \varphi_{(s-2)}^T \cdot \Gamma_{s-2}\\
 &-\frac{m^2}{2} \left[\left(\Phi+\kappa\eta \varphi_{(s-2)}^T\right)^2-\frac{s(s-1)}{2}\left(\Phi^\prime + 2s\kappa\varphi_{(s-2)}^T\right)^2\right] -\left(\Phi+\kappa\eta\varphi_{(s-2)}^T\right) \cdot \tilde{\mathcal{J}}\\
 &+c_{s-1}\left(\phi_{(s-1)}^\prime\right)^2- 4 c_{s-1}\phi_{(s-1)}^\prime \cdot \partial\cdot \varphi_{(s-2)}^T+ a_{1,2}\,\left(\partial\cdot \varphi_{(s-1)}^\prime\right) \cdot \varphi_{(s-2)}^\prime\\
&+c_{s-2}\,\left(\varphi_{(s-2)}^\prime\right)^2 + b_2 \left(\partial_\mu \varphi_{(s-2)}^\prime\right)^2+ \tilde{b}_2\left(\partial\cdot \varphi_{(s-2)}^\prime\right)^2+ \cdots
 \end{split}
\end{align} 
where we have added additional mass terms and kinetic terms of the auxiliary fields.
In what follows, we define the operator
\begin{equation}
\mathcal{I}_k \equiv \frac{m}{\sqrt{s}}\left[2\partial \partial \varphi_{(k)}^T-2\eta\square \varphi_{(k)}^T-\eta \partial \partial\cdot\varphi_{(k)}^T\right]\,,
\end{equation}
which reduces to \eq{defIs2} for $k=s-2$ and is proportional to the variation of the Fronsdal tensor under Weyl-like transformations of the field $\phi_{(k)}$,
\begin{equation}
\phi_{(k)}\rightarrow \phi_{(k)} + \lambda_k \eta \varphi_{(k-2)}^T\,,\qquad \delta \hat{\Gamma}_{(k)} =\lambda_k \frac{\sqrt{s} }{2 m}(k-1)\,\mathcal{I}_{k-2}\,,
\end{equation}
where $\lambda_k$ is the transformation parameter.

In the main text, we have shown that a field redefinition of $\Phi$ (see  Eq.~(\ref{fieldRedefSpinS})) introduces a kinetic term for $\varphi_{(s-2)}^T$ which is invariant only under gauge transformations with transverse gauge parameters. Therefore, the Lagrangian contains ghost-like terms $\sim \left(\partial^2 \varphi_{(s-3)}^T\right)^2$, since the definition
\begin{equation}
\varphi_{s-2}^T=\phi_{(s-2)}^T - \partial\varphi_{(s-3)}^T + \frac{1}{(s-2)}\eta \partial \cdot \varphi_{(s-3)}^T
\end{equation}
does not resemble a transverse gauge transformation. These terms can only be removed if the coefficients $a_{1,2},b_2$ and $\tilde{b}_2$ are tuned to specific values, such that a gauge invariant kinetic term for the traceful field $\varphi_{(s-2)}$ is recovered. For this purpose, we recall that for a massless spin-$k$ field the quadratic Lagrangian can be equivalently written in terms of the traceless fields $\varphi_{(k)}^\prime$ and $ \varphi_{(k)}^T$
\begin{align}
\begin{split}
\mathcal{L}_{(k)} &= -\frac{1}{2}\left(\partial_\mu \varphi_{(k)}^T\right)^2 + \frac{k}{2}\left(\partial \cdot \varphi_{(k)}^T\right)^2 + \frac{(k-1)^2}{2}\varphi_{(k)}^\prime \partial\cdot\partial\cdot \varphi_{(k)}^T \\
&+ \frac{(k-1)^2 (2k-1)}{8k}\left(\partial_\mu \varphi_{(k)}^\prime\right)^2+\frac{(k-1)^2(k-2)^2}{8k}\left(\partial \cdot \varphi_{(k)}^\prime \right)^2.
\end{split}
\end{align}
It is then clear that a gauge invariant kinetic term for $\varphi_{(s-2)}$ is reproduced if we match the coefficients $a_{1,2},b_2,\tilde{b}_2$ with the previous equation
\begin{equation}
\label{tunings}
\begin{split}
a_{1,2} &= -m^2\frac{(2s-1)(s-3)^2}{4}\,,\qquad b_2 = m^2\frac{(2s-1)(s-3)^2(2s-5)}{8(s-2)}\,\\
\tilde{b}_2&= m^2\frac{(2s-1)(s-3)^2(s-4)^2}{8(s-2)}\,,
 \end{split}
\end{equation}
whereas the coefficient $c_{s-2}$ is fixed by demanding a Fronsdal kinetic term for $\varphi_{s-3}^T$, as done previously for $c_{s-1}$. Indeed, let us notice that the transformation Eq.~(\ref{fieldRedefSpinS}) induces a kinetic term for $\varphi_{(s-3)}^T$ through the mass term of $\Phi$, i.e.
\begin{equation}
\label{kinTermMassS}
-\frac{m^4 s (2s-1)(s-2)}{2 (s-1)} \varphi_{(s-3)}^T \Gamma_{s-3}-\frac{m^4 s (2s-1)(2s-5)}{2 (s-1)} \varphi_{(s-3)}^T \cdot \partial \partial \cdot \varphi_{(s-3)}^T\,,
\end{equation}
as well as mixing terms between $\phi_{(s-1)}$ and $\varphi_{(s-3)}^T$
\begin{equation}
\label{mixingSm1Sm3}
\begin{split}
s(1-2s) \kappa \phi_{(s-1)}\cdot \mathcal{I}_{s-3} + \frac{1}{2}m^2(s-3)(2s-1) \phi_{(s-1)}^\prime \cdot \partial \partial \cdot \varphi_{(s-3)}^T\,.
\end{split}
\end{equation}
With the choice of $a_{1,2}$ made in Eq.~(\ref{tunings}), the latter mixing is removed, whereas the former can be removed by a Weyl-like transformation  
\begin{equation}
\phi_{(s-1)}\rightarrow \phi_{(s-1)} + \lambda_{s-1} \eta \varphi_{(s-3)}^T\,,
\end{equation} 
under which the kinetic term of $\phi_{(s-1)}$ transforms as
\begin{align}
\label{varKinSm1}
\delta\left(\phi_{(s-1)}\cdot  \hat\Gamma_{(s-1)}\right)
&=\frac{\sqrt{s}(s-2)}{m}\lambda_{s-1}\phi_{(s-1)}\cdot \mathcal{I}_{s-3} + \lambda_{s-1} \eta \varphi_{(s-3)}^T \cdot \delta \hat{\Gamma}_{(s-1)}\,.
\end{align}
The mixing $\phi_{(s-1)} \cdot \mathcal{I}_{s-3}$  in Eq.~(\ref{mixingSm1Sm3}) then cancels if
\begin{equation}
\lambda_{s-1} = \frac{ m^2(2s-1)}{(s-1)(s-2)}\,.
\end{equation}
Therefore, by summing Eq.~(\ref{kinTermMassS}, \ref{mixingSm1Sm3}, \ref{varKinSm1}) a non-standard kinetic term is generated 
\begin{equation}
\frac{3}{2}m^4 (s-1)(2s-1)\varphi_{(s-3)}^T \cdot \Gamma_{s-3} +m^4(2s-1)(2s-3) \varphi_{(s-3)}^T \cdot \partial \partial\cdot \varphi_{(s-3)}^T\,.
\end{equation}
The coefficient $c_{s-2}$ must be tuned to cancel the last mixing
\begin{equation}
c_{s-2} = m^4 \frac{(s-3)(2s-1)(2s-3)}{4}\,.
\end{equation}

\section{Decoupling limit of massive spin-3 theory}
\label{appendix:decouplingLimit}

In this section, we present a more explicit computation of the decoupling limit of the free massive spin-3 theory. To simplify the computation, we can conveniently choose the gauge $A_\mu=0$. When most of the kinetic mixings with the scalar mode will be removed, we will reintroduce the field $A_\mu$ together with its gauge invariance.

By following our construction in Sec.~\ref{sec:eft}, we consider the Lagrangian in term of the field $\varphi_{\mu\nu} = H_{\mu\nu}+2\partial_\mu\partial_\nu\pi$
\begin{equation}
\mathcal{L}^0=\Phi_{\mu\nu\rho}\hat{\Gamma}_3^{\mu\nu\rho}-\frac{m^2}{2}\left[\Phi_{\mu\nu\rho}^2-3\Phi_\mu^2\right]+\varphi_{\mu\nu}\hat{\Gamma}^{\mu\nu}_2 +\mathcal{L}_\text{mix}^{\Phi \varphi_2}+ \mathcal{L}_{\varphi}^\text{mass} \,,
\end{equation}
where $\varphi_{\mu\nu}\hat{\Gamma}^{\mu\nu}_2$ is the usual linearized Einstein-Hilbert free action and
\begin{align}
\mathcal{L}_\text{mix}^{\Phi \varphi_2} &=\sqrt{3}\,m\left[\Phi_{\mu\nu\rho}\partial^\mu \varphi^{\nu\rho} + \frac{1}{2}\Phi_\mu \partial^\mu \varphi_\alpha^\alpha - 2\Phi_\mu\partial_\nu \varphi^{\mu\nu}\right]\,,\qquad \mathcal{L}_\varphi^\text{mass} = c_1 \left(\varphi_\mu^\mu\right)^2.
\end{align}
The scalar mode $\pi$ enters the definition of $\varphi_{\mu\nu}$ as a gauge transformation and therefore does not affect the Einstein-Hilbert kinetic term. Instead, it affects the mass and mixing terms
\begin{align}
\delta\mathcal{L}_\varphi^\text{mass} &= 4c_1 \left[ H \square \pi+ \left(\square \pi\right)^2\right]\,,\\
\delta \mathcal{L}_\text{mix}^{\Phi \varphi_2} &= \sqrt{3}\,m\left[2\Phi_{\mu\nu\rho}\partial^\mu\partial^\nu\partial^\rho \pi - 3\Phi_\mu \partial^\mu \square \pi \right]\,.\label{mixing30}
\end{align}
We can remove the last mixing kinetic term through the field redefinition
\begin{align}
\label{fieldRedefSpin30}
\Phi_{\mu\nu\rho}&\rightarrow \Phi_{\mu\nu\rho} -\frac{m}{2\sqrt{3}}\, \eta_{(\mu\nu}\partial_{\rho)} \pi\,.
\end{align}
Indeed, this transformation does affect the kinetic term of the spin-3 field
\begin{equation}
\label{transfPhiLagr}
\delta \left(\Phi_{\mu\nu\rho}\hat{\Gamma}_3^{\mu\nu\rho}\right) = \frac{15}{4}m^4 (\partial \pi)^2  + 3m^2\left(\square\pi\right)^2-\sqrt{3}\, m \left(2\Phi^{\mu\nu\rho}\partial_{\mu\nu\rho}\pi-3\Phi_\mu\partial^\mu\square\pi\right) -\frac{5\sqrt{3}}{2} m^3 \phi_\mu\partial^\mu \pi
\end{equation}
providing the term to cancel the mixing between the spin-3 and the scalar Goldstone. Summing also the kinetic terms for $\pi$ generated by the field redefinition Eq.~(\ref{fieldRedefSpin30}) from the spin-3 mass terms and the mixing Eq.~(\ref{mixing30}), we arrive at the Lagrangian
\begin{align}
\mathcal{L}^0&=\Phi_{\mu\nu\rho}\hat{\Gamma}_3^{\mu\nu\rho}-\frac{m^2}{2}\left[\Phi_{\mu\nu\rho}^2-3\Phi_\mu^2\right]+H_{\mu\nu}\hat{\Gamma}^{\mu\nu}_2 +c_1 H^2+\mathcal{L}_\text{mix}^{\Phi H}+\mathcal{L}^{H \pi}_\text{mix} -\frac{5}{2}m^2 \Phi_\mu\partial^\mu\pi \\
&+\frac{15}{4}m^4(\partial \pi)^2  + \left(4c_1-3m^2\right)\left(\square\pi\right)^2 \,,
\end{align}
where 
\begin{align}
\label{kinMixHpi}
\mathcal{L}^{H \pi}_\text{mix} &= \left[\left(4c_1+2m^2\right)\square H-5m^2 \partial_\mu\partial_\nu H^{\mu\nu}\right]\pi\,,\\
\mathcal{L}_\text{mix}^{\Phi H} &=\sqrt{3}\,m\left[\Phi_{\mu\nu\rho}\partial^\mu H^{\nu\rho} + \frac{1}{2}\Phi_\mu \partial^\mu H - 2\Phi_\mu\partial_\nu H^{\mu\nu}\right].
\end{align}
The ghost-like kinetic term $\left(\square \pi\right)^2$ cancels only if $c_1 = 3/4 m^2$. With this choice, the kinetic mixing between $\pi$ and $H^{\mu\nu}$ can be resolved through the field redefinition\footnote{In the Fierz-Pauli Lagrangian, this is the same mixing (up to multiplicative constants) appearing between the spin-2 massive field and the spin-0 Stueckelberg mode.}
\begin{equation}
H_{\mu\nu}\rightarrow H_{\mu\nu} +\frac{5}{2}m^2\,\eta_{\mu\nu} \pi
\end{equation}
under which
\begin{align}
\label{resolvemixingHpi}
\delta \left(\mathcal{L}^{H \pi}_\text{mix}\right) &=-\frac{75}{2} m^4 \left(\partial \pi\right)^2 \,, \qquad 
\delta \left(H_{\mu\nu}\hat{\Gamma}_2^{\mu\nu}\right)= \frac{75}{4}m^4 (\partial \pi)^2 + 5m^2\left[\partial_\mu\partial_\nu H^{\mu\nu}-\square H \right]\pi \,, \\
\delta \left(c_1 H^2\right) &= 15m^4  H \pi + 75 m^6\pi^2 \,, \qquad 
\delta \left( \mathcal{L}^{\Phi H}_\text{mix}\right)=\frac{5\sqrt{3}}{2}m^3\Phi_\mu\partial^\mu \pi\,.
\end{align}
The last term of Eq.~(\ref{resolvemixingHpi}) cancels Eq.~(\ref{kinMixHpi}) if $c_1 = 3/4 m^2$ and the $\Phi_\mu\partial^\mu\pi$ mixing  cancels as well. The final lagrangian is then
\begin{equation}
\label{step1Decoupling}
\begin{split}
\mathcal{L}^0&=\Phi_{\mu\nu\rho}\hat{\Gamma}_3^{\mu\nu\rho}-\frac{m^2}{2}\left[\Phi_{\mu\nu\rho}^2-3\Phi_\mu^2\right]+H_{\mu\nu}\hat{\Gamma}^{\mu\nu}_2 +\frac{3}{4}m^2 H^2+\mathcal{L}_\text{mix}^{\Phi H}+ 15 m^4 H\pi \\
&-15 m^4  \left(\partial \pi\right)^2  + 75 m^6 \pi^2 \,.
\end{split}
\end{equation}
Notice that the tuning of $c_1$ was necessary in order to recover a ghost-free theory. Indeed, if we set $c_1 =\left( \frac{3}{4} + \delta c\right) m^2$ with $\delta c \sim O(1)$ then a ghost appears with a mass $m_\text{ghost}^2 \sim m^2/\delta c $.\\

We can now reintroduce the vector modes $A_\mu$ by redefining $H_{\mu\nu}\rightarrow H_{\mu\nu} - \partial_{(\mu}A_{\nu)}$. The computation is now  simpler as most of the mixing terms have been removed. From our result Eq.~(\ref{step1Decoupling}), it is clear that $A_\mu$ will mix with $\Phi_{\mu\nu\rho}$ and $\pi$. 
Once we reintroduce the vector modes, we have the following additional terms 
\begin{align}
\delta\left( \mathcal{L}_\text{mix}^{\Phi H}\right) &=\mathcal{L}_\text{mix}^{\Phi A} \equiv -\sqrt{3}\,m \left[2\Phi_{\mu\nu\rho}\partial^\mu\partial^\nu A^\rho - \Phi_\mu \partial^\mu \partial^\nu A_\nu - 2\Phi_\mu \square A^\mu\right] \,, \label{mixing31} \\
 \label{changeMassH}
\delta\left(\frac{3}{4}m^2 H^2\right) &=3m^2\left(\partial_\mu A^\mu\right)^2-3m^2\partial_\mu A^\mu H\,,\qquad \delta\left(15 m^4 H\pi\right)  = -30 m^4\partial_\mu A^\mu \pi \,.
\end{align}
The kinetic mixing between $\Phi_{\mu\nu\rho}$ and $A^\mu$ in Eq.~(\ref{mixing31}) can be removed with the following field redefinition
\begin{equation}
\Phi_{\mu\nu\rho} \rightarrow \Phi_{\mu\nu\rho} +\frac{1}{2\sqrt{3}}m\, \eta_{(\mu\nu}A_{\rho)}\,,
\end{equation}
under which
\begin{align}
\label{TransKinTerm3}
\delta \left(\Phi_{\mu\nu\rho}\hat{\Gamma}_3^{\mu\nu\rho}\right) &=-\mathcal{L}_\text{mix}^{\Phi A} + \frac{m^2}{2}\left[\left(\partial_\mu A^\mu\right)^2 + 5 \left(\partial_\mu A_\nu\right)^2\right] \,, \\
\delta\left(-\frac{m^2}{2}\left[\Phi_{\mu\nu\rho}^2-3\Phi_\mu^2\right]\right)&= \frac{15}{4}m^4 A_\mu^2 +\frac{5}{2} m^2 A_\mu\phi^\mu \, \\
\delta \left(\mathcal{L}_\text{mix}^{\Phi A}\right) &=-m^2\left[5\left(\partial_\nu A_\mu\right)^2+\left(\partial_\mu A^\mu\right)^2\right] \,, \\
\delta \left(\mathcal{L}_\text{mix}^{\Phi H}\right) &= \mathcal{L}_\text{mix}^{A H}\equiv m^2  \left[2A_\mu \partial^\mu H -5A_\mu \partial_\nu H^{\mu\nu}\right] \,.
\end{align}
The mixing Eq.~(\ref{mixing31}) cancels with the first term of Eq.~(\ref{TransKinTerm3}) and a gauge invariant kinetic term for the vector modes is generated
\begin{align}
-\frac{5}{4}m^2 F_{\mu\nu}^2\,.
\end{align}
Summing all the terms we get the Lagrangian
\begin{equation}
\begin{split}
\mathcal{L}^{0} &=\Phi_{\mu\nu\rho}\hat{\Gamma}_3^{\mu\nu\rho}+H_{\mu\nu}\hat{\Gamma}^{\mu\nu}_2 -\frac{5}{4}m^2 F_{\mu\nu}^2 -15m^4 \left(\partial \pi\right)^2  \\
&-\frac{m^2}{2}\left[\Phi_{\mu\nu\rho}^2-3\Phi_\mu^2\right]+\frac{3}{4}m^2 H^2 + \frac{15}{4}m^4 A_\mu^2+ 75 m^6\pi^2\\
&+\mathcal{L}_\text{mix}^{\Phi H}+ \mathcal{L}_\text{mix}^{AH} + 15 m^4 \left[H\pi  - 2\partial_\mu A^\mu \pi\right] +\frac{5}{2}m^2 A_\mu\phi^\mu-3m^2  \partial_\mu A^\mu H \,,
\end{split}
\end{equation}
which is smooth in the limit $m\rightarrow 0$, once all the fields are canonically normalized.

\section{Tuned potential for arbitrary spins}
\label{appendix:tunedPotential}

In this Appendix, we show that our understanding of the mixings between the transverse and longitudinal modes allows us to explicitly realize the best energy growth of four-scalar amplitudes from zero-derivative interactions, namely $\mathcal{M}\sim E^{3s}$ and $\mathcal{M}\sim E^{3s+1}$ for even and odd spin respectively, which was conjectured in Ref.~\cite{Bonifacio:2018vzv}. The potential that gives rise to such a behavior is a straightforward generalization of Eq.~(\ref{tunedPotential}), which can be conveniently written as
\begin{align}
\label{tunedPotentialHS}
s \text{ even:}\,\qquad &\epsilon^{\mu_1...\mu_4}\epsilon^{\nu_1...\nu_4}\cdots\epsilon^{\rho_1...\rho_4}\, \Phi_{\mu_1\nu_1...\rho_1}\Phi_{\mu_2\nu_2...\rho_2}\Phi_{\mu_3\nu_3...\rho_3}\Phi_{\mu_4\nu_4...\rho_4}\,,\\
s \text{ odd:}\,\qquad &\epsilon^{\mu_1...\mu_4}\epsilon^{\nu_1...\nu_4}\cdots\epsilon^{\rho_1...\rho_4} \Phi_{\mu_1\nu_1...\rho_1\alpha}\Phi_{\mu_2\nu_2...\rho_2}^{\phantom{\mu_2\nu_2...\rho_2}\alpha}\Phi_{\mu_3\nu_3...\rho_3\beta}\Phi_{\mu_4\nu_4...\rho_4}^{\phantom{\mu_2\nu_2...\rho_2}\beta} \,,
\end{align}
and consists of $s$ and $s-1$ $\epsilon$-tensors for even and odd spin respectively. The scalar interactions in terms of the Stueckelberg scalar field $\pi\equiv \phi_{(0)}$ can be read through the gauge invariant combination $D{\varphi_{(s-2)}}$ defined in Eq.~(\ref{generalCovD}) away from the unitary gauge. Generically, the leading interaction is of the form $\left(\partial^s \pi\right)^4$, but the specific contractions with the $\epsilon$-tensors in Eq.~(\ref{tunedPotentialHS}) makes this term vanish up to total derivatives. The would-be subleading terms now dominate the amplitude and originate from non-vanishing terms proportional to
\begin{align}
\label{leadTerm}
s \text{ even:}\,\qquad &\epsilon^{\mu_1...\mu_4}\epsilon^{\nu_1...\nu_4}\cdots\epsilon^{\rho_1...\rho_4} \partial_{\mu_1\nu_1...\rho_1}\pi\,\partial_{\mu_2\nu_2...\rho_2}\pi\,\partial_{\mu_3\nu_3...\rho_3}\pi\, D{\varphi_{(s-2)}}_{\mu_4\nu_4...\rho_4}\,,\\
s \text{ odd:}\,\qquad &\epsilon^{\mu_1...\mu_4}\epsilon^{\nu_1...\nu_4}\cdots\epsilon^{\rho_1...\rho_4} \partial_{\mu_1\nu_1...\rho_1\alpha}\pi\,\partial_{\mu_2\nu_2...\rho_2}^{\phantom{\mu_2\nu_2...\rho_2}\alpha}\pi\,\partial_{\mu_3\nu_3...\rho_3\beta}\pi\, D{\varphi_{(s-2)}}_{\mu_4\nu_4...\rho_4}^{\phantom{\mu_2\nu_2...\rho_2}\beta}.
\end{align}
In this basis, the Stueckelberg field $\pi$ has not a proper kinetic term, which must be induced by resolving the mixings of the transverse and longitudinal modes. As we have shown in Sec.~\ref{sec:eft} and Appendix~\ref{appendix:tuning} this is done via generalized Weyl transformations
\begin{align}
\label{WeylTransfGenHS}
\begin{split}
\Phi_s &\rightarrow \Phi_s + \kappa\,\eta\, \phi_{(s-2)}\\
\phi_{(s-1)} &\rightarrow \phi_{(s-1)}  + \lambda_{s-1}\eta\, \phi_{(s-3)}\\
\phi_{(s-2)} &\rightarrow \phi_{(s-2)}  + \lambda_{s-2}\eta\, \phi_{(s-4)}\\
\dots \\
\phi_{(2)} &\rightarrow \phi_{(2)}  + \lambda_{2}\eta\, \pi
\end{split}
\end{align}
where $\kappa, \lambda_k$ are chosen such as to generate gauge-invariant kinetic terms for the Goldstone fields. Their value in unimportant for the purpose of this discussion. 

Let us focus on the even and odd spins cases separately.
\paragraph*{Even spin.}
The leading scalar interaction is obtained by applying the cascade of transformations in Eq.~(\ref{WeylTransfGenHS}) to the non-vanishing term Eq.~(\ref{leadTerm}). For even spins, the scalar field is obtained as the result of the chain of transformations $\Phi_s\rightarrow\phi_{(s-2)}\rightarrow \cdots \rightarrow\pi$ under which
\begin{equation}
\epsilon^{\mu_1...\mu_4}\epsilon^{\nu_1...\nu_4}\cdots\epsilon^{\sigma_1...\sigma_4}\epsilon^{\rho_1...\rho_4}\, \partial_{\mu_1\nu_1...\sigma_1\rho_1}\pi\,\partial_{\mu_2\nu_2...\sigma_2\rho_2}\pi\,\partial_{\mu_3\nu_3...\sigma_3\rho_3}\pi\,\eta_{(\mu_4\nu_4}...\eta_{\sigma_4\rho_4)}\pi
\end{equation}
is generated. This term is not vanishing and leads to a scalar amplitude $\mathcal{M}\sim E^{3s}$, as clearly seen by counting derivatives.
\paragraph*{Odd spin.} In this case, the scalar field comes from the chain of transformations starting from the spin-$(s-1)$ Goldstone field $\phi_{(s-1)}\rightarrow\phi_{(s-3)}\rightarrow...\rightarrow\pi$. The four-scalar interaction that is then generated is
\begin{equation}
\epsilon^{\mu_1...\mu_4}\epsilon^{\nu_1...\nu_4}\cdots\epsilon^{\sigma_1...\sigma_4}\epsilon^{\rho_1...\rho_4}\, \partial_{\mu_1\nu_1...\sigma_1\rho_1\alpha}\pi\,\partial_{\mu_2\nu_2...\sigma_2\rho_2}^{\phantom{\mu_2\nu_2...\sigma_2\rho_2}\alpha}\pi\,\partial_{\mu_3\nu_3...\sigma_3\rho_3}^{\phantom{\mu_3\nu_3...\sigma_3\rho_3}\beta}\pi\,\eta_{(\mu_4\nu_4}...\eta_{\sigma_4\rho_4}\partial_{\beta)}\pi \,.
\end{equation}
Upon symmetrization and integration by parts, the only non-vanishing term is the one where the index $\beta$ lies on the derivative, which leads to $\mathcal{M}\sim E^{3s+1}$.\\

Incidentally, these interactions are symmetric under the polynomial shifts Eq.~(\ref{LevelN}) with $N=s-1$.

\end{appendices}

\bibliographystyle{utphys}
\bibliography{Bib2} 
\end{document}